\documentclass[a4paper]{article} 
\usepackage[fleqn]{amsmath}
\usepackage{amsfonts,amssymb}
\usepackage{pifont}
\usepackage{mathrsfs}
\usepackage{graphicx}
\usepackage{savesym} 
\savesymbol{AND} 
\usepackage{algorithmic}
\usepackage{algorithm}
\usepackage{accents}
\usepackage{bm}
\usepackage{multirow}
\usepackage{paralist}
\usepackage{caption}
\usepackage{subcaption}
\usepackage[%
    all,
    defaultlines=3 
]{nowidow}

\newcounter{Thmnbr}
\def\theenumi{{\arabic{enumi}}.}     
\def\theenumii{\bff{\roman{enumii}}.}       
     \def\theenumiii{\alph{enumiii}}
       
\def\p@enumii{\theenumi}
\def\p@enumiii{\theenumi(\theenumii)}
\def\p@enumiv{\p@enumiii\theenumiii}

\newtheorem{lemm}{Lemma}[section]
\newtheorem{theo}[lemm]{Theorem}
\newtheorem{propo}[lemm]{Proposition}
\newtheorem{coro}[lemm]{Corollary}
\newtheorem{assump}[lemm]{Assumptions}
\newtheorem{aim}[lemm]{Aims}

\newtheorem{defi}{Definition}[section]
\newtheorem{rema}{Remark}[section]

\newcommand{\findemo}{{\hfill \ding{113}}}

\newcommand{\debdemo}{{\noindent {\bf  Proof}.\  }}

\newcommand{\R}{{\mathop{\rm I\mkern -3.5mu R}}}
\newcommand{\N}{{\mathop{\rm I\mkern -3.5mu N}}}

\newcommand{\Rn}{{{\R}^n}}
\newcommand{\Rm}{{{\R}^m}}

\newcommand{\Rnn}{{{\R}^{n\times n}}}

\newcommand{\Rnm}{{{\R}^{n\times m}}}

\newcommand{\Rnl}{{{\R}^{n\times l}}}

\newcommand{\ith}{$^{\textrm{th}}$\ }

\newcommand{\cf}{{\it cf.\ }} 

\newcommand{\ssi}{{\emph{if and only if\ }}}
\newcommand{\ie}{{\it i.e.}} 
\newcommand{\tq}{\emph{s.t.}}
\newcommand{\wrt}{\emph{w.r.t. }}
\newcommand{\resp}{\emph{resp}}

\let\d\displaystyle
\let\nn\nonumber

\newcommand{\nid}{{\noindent}}

\DeclareMathOperator*{\diag}{Diag}

\DeclareMathOperator*{\bdiag}{Bdiag}
\DeclareMathOperator*{\tr}{tr} 

\DeclareMathOperator*{\rank}{rank}

\DeclareMathOperator*{\dif}{d\!}
\DeclareMathOperator*{\card}{Card}
\DeclareMathOperator*{\mean}{mean}

\newcommand{\norm}[1]{\left\Vert#1\right\Vert}

\newcommand{\abs}[1]{{\ensuremath{\left| #1 \right|}}}
\newcommand{\bff}[1]{\mbox{\boldmath$ #1 $}} 
\newcommand{\eqd}{{\ensuremath{\, =: \,}}}
\newcommand{\eqg}{{\ensuremath{\, := \,}}}
\newcommand{\eq}{\ensuremath{\Leftrightarrow}}
\newcommand{\guil}[1]{{``#1''}}

\newcommand{\BC}{{\mathcal{B}}}
\newcommand{\CC}{{\mathcal{C}}}
\newcommand{\DC}{{\mathcal{D}}}
\newcommand{\EC}{{\mathcal{E}}}

\newcommand{\GC}{{\mathcal{G}}}
\newcommand{\HC}{{\mathcal{H}}}

\newcommand{\KC}{{\mathcal{K}}}

\newcommand{\NC}{{\mathcal{N}}}
\newcommand{\OC}{{\mathcal{O}}}

\newcommand{\PC}{{\mathcal{P}}}
\newcommand{\SC}{{\mathcal{S}}}

\newcommand{\ZC}{{\mathcal{Z}}}

\newcommand{\Fd}{{\mathscr{F}}}

\def\u{\bff{u}}

\def\x{\bff{x}}
\def\y{\bff{y}}
\def\zb{\bff{z}}

\def\taub{\bff{\tau}}

\def\r{\bff{r}}

\def\bdel{\bar{\bff{\delta}}}

\def\bybk{\bar{\y}_{k}}

\def\Akk{A_{k-1}}

\def\Rkk{R_{k-1}}
\def\Rk{R_{k}}

\def\taubk{\bff{\tau}_k}
\def\taubkk{\bff{\tau}_{k-1}}

\def\rki{\bff{r}_{{k}_i}}
\def\rkit{\bff{r}_{{k}_i}^T}
\def\rkki{\bff{r}_{{k-1}_i}}
\def\rkkit{\bff{r}_{{k-1}_i}^{T}}

\def\yk{\bff{y}_{k}} 
\def\delk{\bff{\delta}_{k}} 
\def\vk{\bff{v}_{k}}
\def\wk{\bff{w}_{k}}
\def\wkk{\bff{w}_{k-1}} 

\def\xe{\bff{\hat{x}}}

\def\xz{\bff{\hat{x}_{0}}} 
\def\xza{\bff{x_{0}}} 
\def\xk{\hat{\bff{x}}_{k}} 
\def\xkk{\hat{\bff{x}}_{k/k-1}} 

\def\xkkk{\hat{\bff{x}}_{k-1}} 
\def\dxk{\tilde{\bff{x}}_{k}} 
\def\dxkk{\tilde{\bff{x}}_{k/k-1}} 
\def\dxkkk{\tilde{\bff{x}}_{k-1}} 
\def\xka{\bff{x}_{k}} 
\def\xkka{\bff{x}_{k-1}} 
\def\bxk{\breve{\bff{x}}_{k}}

\def\zbz{\bff{z}_0}

\def\delkT{\bff{\delta}_{k}^T}

\def\Vscr{\mathscr{V}}

\def\Pk{P_{k}}
\def\Pki{P_{k_{i}}}

\def\Pkiii{P_{k_{i-1}}}
\def\Pkz{P_{k_{0}}}

\def\Pkk{P_{k/k-1}}
\def\Pkki{P_{{k/k-1}_{i}}}

\def\Pkkz{P_{{k/k-1}_{0}}}

\def\Pkkk{P_{k-1}}

\def\sigz{\varsigma_{0}}

\def\sigk{\varsigma_{k}}

\def\sigkkk{\varsigma_{k-1}}

\def\ki{_{k_i}}

\def\kk{_{k/k-1}}

\def\kkk{_{k-1}}

\def\zero{{\mathbf{0}}} 

\def\kkp{_{{k-1}_+}}

\def\yki{\bff{y}_{k_i}}

\def\Rpi{{\R}^{p_i}}

\def\xki{\hat{\bff{x}}_{k_i}}

\def\xkiii{\hat{\bff{x}}_{k_{i-1}}}
\def\xkz{\hat{\bff{x}}_{k_{0}}}

\def\gamki{\gamma_{k_i}}

\def\yki{{y_{k_i}}}
\def\byki{\bar{y}_{k_i}}

\def\sigki{\varsigma_{k_i}}

\def\sigkiii{\varsigma_{k_{i-1}}}
\def\sigkz{\varsigma_{k_0}}

\def\delki{{\delta_{k_i}}}
\def\delkis{{\delta_{k_i}^{2}}}

\def\omki{\omega_{k_i}}

\DeclareMathOperator*{\mink}{\oplus}

\makeatletter
\DeclareRobustCommand\sfrac[1]{\@ifnextchar/{\@sfrac{#1}}%
                                            {\@sfrac{#1}/}}
\def\@sfrac#1/#2{\leavevmode\kern.1em\raise.5ex
         \hbox{$\m@th\fontsize\sf@size\z@
                           \selectfont#1$}\kern-.1em
         /\kern-.15em\lower.25ex
          \hbox{$\m@th\fontsize\sf@size\z@
                            \selectfont#2$}}
\makeatother
\makeatletter
\newcommand{\mathl}{\@fleqntrue\@mathmargin0pt} 
\newcommand{\mathc}{\@fleqnfalse} 
\newcommand{\pushr}[1]{\ifmeasuring@#1\else\omit\hfill$\displaystyle#1$\fi\ignorespaces}
\newcommand{\pushl}[1]{\ifmeasuring@#1\else\omit$\displaystyle#1$\hfill\fi\ignorespaces}
\makeatother

\DeclareMathAlphabet{\mathpzc}{OT1}{pzc}{m}{it}                                                        
\allowdisplaybreaks

\makeatletter
\newcounter{ALC@tempcntr}
\newcommand{\LCOMMENT}[1]{%
    \setcounter{ALC@tempcntr}{\arabic{ALC@rem}}
    \setcounter{ALC@rem}{1}
    \item \{#1\}
    \setcounter{ALC@rem}{\arabic{ALC@tempcntr}}
}%
                                                          
\begin{document}

\title{Ellipsoidal  constrained state estimation in presence of bounded disturbances}
\author{Yasmina BECIS-AUBRY
\thanks{*The author is with Universit\'{e} d'Orl\'{e}ans, Laboratoire PRISME EA 4229 (Univ. Orl\'{e}ans - INSA CVL). 63 av. de Lattre de Tassigny, 18020 Bourges Cedex, FRANCE. Tel. +33 2 48 23 84 78 {\tt\small Yasmina.Becis@univ-orleans.fr}.}
}


\maketitle

\begin{abstract}

This contribution proposes a recursive, computationally efficient, ready-to-use, online method for the ellipsoidal state characterization for linear discrete-time models with additive unknown disturbances vectors (bounded by known possibly degenerate zonotopes) corrupting both the state difference equation and the sporadic measurement vectors, which are expressed as linear inequality and equality constraints on the state vector.

The algorithm is decomposed into {\it time updating} and {\it observation updating} steps. In the latter, a suitable switching estimation gain is designed in such a way as to ensure the input-to-state stability of the estimation error.

%

\end{abstract}


\setlength{\abovedisplayskip}{0pt}
\setlength{\belowdisplayskip}{0pt}
\setlength{\abovedisplayshortskip}{0pt}
\setlength{\belowdisplayshortskip}{0pt}

\newlength{\mes}
\setlength{\mes}{5em}
\newcommand{\underb}[1]{\underaccent{\bar}{#1}}
\newcommand{\Rrho}{{{\R}^{\varrho}}}
\newcommand{\Rrhoq}{{{\R}^{\varrho\times q}}}
\newcommand{\Rqrho}{{{\R}^{\varrho\times q}}}
\newcommand{\Rrhorho}{{{\R}^{\varrho\times\varrho}}}
\def\Rpi{\R^\pi}
\def\Rnpi{\R^{n\times\pi}}
\def\Rnrho{\R^{n\times\varrho}}
\def\txt{\textstyle}
\def\sqrtexp{^{\frac{1}{2}}}
\newcommand{\kkj}[1]{_{{k/k-1}_{#1}}}
\def\kkp{_{k+1/k}}
\newcommand{\kkpj}[1]{_{{k+1/k}_{#1}}}
\def\ybk{\bm{y}_{k}}
\def\xibk{\bm{\xi}_k}
\def\xibkk{\bm{\xi}_{k-1}}
\def\xibh{\hat{\bm{\xi}}}
\def\xibhk{\hat{\bm{\xi}}_{k}}
\def\xibhkk{\hat{\bm{\xi}}_{k/k-1}}
\def\xibhkkk{\hat{\bm{\xi}}_{k-1}}
\def\etabk{\bm{\eta}_k}
\def\etabkk{\bm{\eta}_{k-1}}
\def\dbk{\bm{d}_k}
\def\dbt{\bm{d}^T}
\newcommand{\bk}[1]{{b}_{k_{#1}}}
\newcommand{\bki}{{b}_{k_{i}}}
\def\udel{\underb{\delta}}
\def\bdel{\bar{\delta}}
\def\udelkis{\underb{\delta}_{k_i}^{2}}
\def\udelkj{\underb{\delta}_{k_j}}
\def\uyki{\underb{y}_{k_i}}
\def\uybk{\underb{\bm{y}}_{k}}
\def\bybk{\bar{\bm{y}}_{k}}
\def\bbybk{\bar{\bar{\bm{y}}}_{k}}
\def\bbyk{\bar{\bar{{y}}}_{k}}
\def\yb{\bm{y}}
\def\xb{\bm{x}}
\def\ab{\bm{a}}
\def\bb{\bm{b}}
\def\cb{\bm{c}}
\def\db{\bm{d}}
\def\gb{\bm{g}}
\def\jb{\bm{j}}
\def\lb{\bm{l}}
\def\nb{\bm{n}}
\def\qb{\bm{q}}
\def\bgb{\bar{\bm{g}}}
\def\ub{\bm{u}}
\def\ubk{\ub_{k}}
\def\mub{\bm{\mu}}
\def\mubk{\bm{\mu}_k}
\newcommand{\muki}[1]{{\mu}_{k_{#1}}}
\def\omb{\bm{\omega}}
\def\bomb{\bar{\bm{\omega}}}
\def\ombk{\bm{\omega}_k}
\def\bombk{\bomb_k}
\def\nub{\bm{\nu}}
\def\nubk{\bm{\nu}_k}
\def\nubkk{\bm{\nu}_{k-1}}
\def\bnubk{\bar{\bm{\nu}}_k}
\def\bnuki{\bar{\nu}_{k_i}}

\def\bPsi{\bar{\Psi}}
\def\bPsik{\bar{\Psi}_k}
\def\psib{\bm{\psi}}
\def\bpsib{\bar{\psib}}
\def\bpsibk{\bar{\psib}_k}
\def\bZk{\bar{\ZC}_k}
\def\bEk{\bar{\EC}_k}
\def\bPk{\bar{P}_k}
\def\bPhik{\bar{\Phi}_k}
\def\bOmk{\bar{\Omega}_k}

\def\bsigk{\bar{\varsigma}_k}
\def\bxk{\bar{\bm{x}}_k}

\def\ccb{\check{\cb}}
\def\bccb{\bar{\ccb}}
\def\bbccb{\bar{\bccb}}
\def\bcb{\bar{\cb}}
\def\bbcb{\bar{\bcb}}
\def\eb{\bm{e}}
\def\hb{\bm{h}}
\def\bxb{\bar{\xb}}
\def\bbxb{\bar{\bxb}}
\def\dxbk{\tilde{\xb}_k}
\def\dbxbk{\tilde{\bxb}_k}
\newcommand{\dbxbki}[1]{\tilde{\bxb}_{k_{#1}}}
\newcommand{\dxbki}[1]{\tilde{\xb}_{k_{#1}}}
\def\by{\bar{y}}
\def\uy{\underb{y}}
\def\bby{\bar{\by}}
\def\fb{\bm{f}} 
\def\bfb{\bar{\fb}} 
\def\bfk{\bar{\fb}}
\def\bbfb{\bar{\bfb}}
\def\bP{\bar{P}}
\def\bbP{\bar{\bP}}
\def\cP{\check{P}}
\def\bcP{\bar{\cP}}
\def\bbcP{\bar{\bcP}}
\def\csig{\check{\sigma}}
\def\uFk{\underb{F}_{k}}
\newcommand{\brFk}{\breve{F}_{k}}
\newcommand{\fbki}{\bm{f}_{k_{i}}}
\newcommand{\fbk}[1]{\bm{f}_{k_{#1}}}
\newcommand{\bfbk}[1]{\bar{\bm{f}}_{k_{#1}}}
\newcommand{\bfbki}{\bar{\bm{f}}_{k_{i}}}
\newcommand{\brfbk}[1]{\breve{\bm{f}}_{k_{#1}}}
\newcommand{\cfbk}[1]{\check{\bm{f}}_{k_{#1}}}
\newcommand{\ufbk}[1]{\underb{\bm{f}}_{k_{#1}}}
\newcommand{\uyk}[1]{\underb{{y}}_{k_{#1}}}
\newcommand{\udek}[1]{\underb{\delta}_{k_{#1}}}
\newcommand{\udeki}{\underb{\delta}_{k_{i}}}
\newcommand{\bdek}[1]{\bar{\delta}_{k_{#1}}}
\newcommand{\bdeki}{\bar{\delta}_{k_{i}}}
\newcommand{\urho}{\underb{\rho}}
\newcommand{\brho}{\bar{\rho}}
\newcommand{\urhok}[1]{\underb{{\rho}}_{k_{#1}}}
\newcommand{\urhoki}{\underb{{\rho}}_{k_{i}}}
\newcommand{\brhok}[1]{\bar{{\rho}}_{k_{#1}}}
\newcommand{\brhoki}{\bar{{\rho}}_{k_{i}}}
\def\uybk{\underb{\y}_{k}}
\newcommand{\byk}[1]{\bar{{y}}_{k_{#1}}}
\newcommand{\brybk}{\breve{\y}_{k}}
\newcommand{\baki}[1]{\bar{a}_{k_{#1}}}
\newcommand{\bomi}[1]{\bar{\omega}_{#1}}
\newcommand{\bomki}[1]{\bar{\omega}_{k_{#1}}}
\newcommand{\ak}[1]{{a}_{k_{#1}}}
\newcommand{\aki}{{a}_{k_{i}}}
\newcommand{\caki}[1]{\check{a}_{k_{#1}}}
\newcommand{\cyk}[1]{\check{y}_{k_{#1}}}
\newcommand{\cybk}{\check{\yb}_{k}}
\newcommand{\bvk}[1]{\bar{\bm{v}}_{k_{#1}}}
\newcommand{\gbkj}[1]{{\bm{g}}_{k_{#1}}}
\newcommand{\bgbkj}[1]{\bar{\bm{g}}_{k_{#1}}}
\newcommand{\bdbki}[1]{\bar{\bm{d}}_{k_{#1}}}
\newcommand{\cdbki}[1]{\check{\bm{d}}_{k_{#1}}}
\newcommand{\dbki}[1]{{\bm{d}}_{k_{#1}}}
\newcommand{\rbki}[1]{{\bm{r}}_{k_{#1}}}
\newcommand{\vkj}[1]{{\bm{v}}_{k_{#1}}}
\newcommand{\zkj}[1]{{z}_{k_{#1}}}
\newcommand{\bzkj}[1]{\bar{z}_{k_{#1}}}
\newcommand{\ukj}[1]{{\bm{u}}_{k_{#1}}}
\newcommand{\alphak}[1]{\alpha_{k_{#1}}}
\newcommand{\betak}[1]{\beta_{k_{#1}}}
\newcommand{\gammak}[1]{\gamma_{k_{#1}}}
\newcommand{\alphaki}{\alpha_{k_{i}}}
\newcommand{\betaki}{\beta_{k_{i}}}
\newcommand{\gammaki}{\gamma_{k_{i}}}
\newcommand{\lamki}{\lambda_{k_{i}}}
\newcommand{\lamk}[1]{\lambda_{k_{#1}}}
\newcommand{\thetaki}{\theta_{k_{i}}}
\newcommand{\alphakinv}[1]{\alpha_{k_{#1}}^{-1}}
\newcommand{\balphak}[1]{\bar{\alpha}_{k_{#1}}}
\newcommand{\balphakinv}[1]{\bar{\alpha}_{k_{#1}}^{-1}}
\newcommand{\bbFk}{\bar{\bar{F}}_k}
\newcommand{\phibki}{{\bm{\varphi}}_{k_{i}}}
\newcommand{\phibk}[1]{{\bm{\varphi}}_{k_{#1}}}
\newcommand{\psibk}[1]{{\psib}_{k_{#1}}}
\newcommand{\bpsibki}[1]{{\bpsib}_{k_{#1}}}
\newcommand{\xbki}[1]{{\bm{\hat{x}}}_{k_{#1}}}
\newcommand{\bpk}{\bar{p}_{k}}
\newcommand{\racine}[1]{{#1}^{\frac{1}{2}}}
\newcommand{\iracine}[1]{{#1}^{-\frac{1}{2}}}
\newcommand{\iracineT}[1]{{#1}^{-\frac{T}{2}}}
\newcommand{\racineT}[1]{{#1}^{\frac{T}{2}}}
\newcommand{\ppzc}{\mathpzc{P}}
\newcommand{\zpzc}{\mathpzc{Z}}
\newcommand{\gpzc}{\mathpzc{G}}
\newcommand{\range}{\mathpzc{R}}
\newcommand{\nul}{\mathpzc{Ker}}
\newcommand{\kf}{\mathscr{K}}
\newcommand{\pscrk}{\mathscr{P}_k}
\newcommand{\zscrk}{\mathscr{Z}_k}
\newcommand{\gscrk}{\mathscr{G}_k}
\newcommand{\ugscrk}{\underb{\mathscr{G}}_k}
\newcommand{\bgscrk}{\bar{\mathscr{G}}_k}
\newcommand{\mscrk}{\mathscr{M}_k}
\newcommand{\hscr}[1]{\mathscr{H}_{#1}}
\newcommand{\hscrk}{\mathscr{H}_k}
\newcommand{\dscrk}{\mathscr{D}_k}
\newcommand{\bGC}{\bar{\GC}}
\newcommand{\uGC}{\underb{\GC}}

\setlength{\mathindent}{0pt}

\section{INTRODUCTION}\label{sec_intro}
There is no more need to praise the interests of the set-membership state estimation techniques neither is there a necessity to recall how interesting alternative they offer to conventional state estimation methods where the statistical assumptions on the disturbances can not be satisfied in certain practical situations, nor how increasing attention they are currently receiving since the noises of their models are assumed only to be bounded.
Nevertheless, the stability question is rarely addressed in this kind of estimation approach. 
\newline\indent
On the other hand, the constrained state filtering has been widely studied in stochastic context
\cite{sim:10}, \cite{dua:13}, \cite{jia:13}.
In \cite{And:19}, constrained Kalman filter variations were reexamined and an alternative derivation of the optimal constrained Kalman filter for time variant systems was proposed.
%
The literature is less abundant on this subject when it comes to bounded error framework. LMI techniques were employed for ellipsoidal set-membership constrained state filtering with linear \cite{yan:09} and linearized nonlinear \cite{yan:09b} equalities. 
In \cite{noa:15}, the authors used a combined stochastic and set-membership uncertainty representation by integrating, into the Kalman filter structure, ellipsoidal constraints on the state vector as a 
relaxation of equality constraints. All these works faced a same challenge, not arising here, in inverting the estimation error covariance matrix, becoming inevitably singular, when dealing with equality constraints.

In our early paper \cite{Bec:08}, we presented a state bounding estimation algorithm for linear discrete-time systems, where the state and all the (process and measurement) disturbances were characterized by multidimensional ellipsoids. 
In order to guarantee  the input-to-state stability of the estimation error and the size decrease of the state bounding ellipsoid at the measurement updating stage, a polynomial equation had to be solved, at each time step, involving the computation of the eigenvalues and eigenvectors of a matrix having the same dimension as the state vector.
This issue was partially solved in \cite{She:18} by overbounding, by a parallelotope, the output noises in the measurement correction step initially characterized by an ellipsoid. Even if this approach was computationally very attractive, it was somehow conservative, because the output disturbances were overbounded twice: first by an ellipsoid then by a parallelotope.
There also remained a non linear equation to solve at the time prediction step (while overbounding geometric sum of two ellipsoids) 
whenever the volume, rather than the squared axes sum of the resulting ellipsoid was to be minimized. 
\newline\indent
Looking more generally into the set-membership techniques, they can mainly be separated in two families: 1.~those using the bound on the $2-$norm of some quantities of interest, resulting typically in ellipsoidal bounding sets that are nothing else that balls or hyperspheres undergoing rotations and scalings and 2.~those  bounding the $\infty-$norm of such quantities, leading to intervals (which are boxes or hypercubes), parallelotopes (skewed, stretched, or shrunken images of such boxes) or zonotopes (flattened images of boxes, which are also generalization of parallelotopes). 
The drawbacks of using exclusively the ellipsoid as bounding set for both state and disturbances vectors were highlighted above. Now, when using intervals, the recourse to interval computing softwares including time costly operations such as subpavings and contractors are inevitable to overcome the conservatism of such aligned with the coordinates axes boxes \cite{Jau:12}, \cite{Ram:15}.
And when dealing with zonotopes to characterize the outer bound of the set of all possible values of the state vector, it is necessary to use some tools such as LMI to manage the growth of the number of generators, inherent to the zonotopes summing, during the time update, and to their intersection, during the measurement correction (\cf \cite{Com:15} and references within).
This is why we chose two different bounding techniques: the ellipsoids (based on the $2-$norm) to characterize the set of all possible values of the state vector at each time step and the zonotope (based on the $\infty-$norm)  to bound the disturbances. Moreover, we are interested here in the state estimation of linear discrete-time systems subject not only to bounded process and measurement disturbances but also to all kinds of linear constraints applied to the state vector, \ie,  equalities (modelled by hyperplanes) and inequalities (represented by polyhedrons and zonotopes); all, noises and constraints, manifesting themselves sporadically, 
not at all time steps. 
\newline\indent The paper is organized as follows. After this introduction, which is completed by some notations and definitions, 
in the second section, the constrained set-membership state estimation problem with sporadic measurements is formulated. The third section concerns the time-prediction stage, while the correction stage of the estimation algorithm is detailed in forth one.  Its properties and stability are studied in the fifth section. Numerical simulations are presented in the sixth and finally, a brief conclusion terminates the paper.
\vspace{8pt}
\newline
{\bf Notations and definitions}
{%
\begin{enumerate}
\item The symbol $\eqg$ (\resp. $\eqd$) means that the {Left Hand Side} (\resp.   {RHS}) \emph{is defined to be equal to} the {Right Hand Side} (\resp.   {LHS}). 
Normal lowercase letters are used for scalars, capital letters for matrices, bold lowercase letters for vectors and calligraphic capital letters for sets.
$\R$, $\R^*$, $R_+$,  $R_+^*$ denote the sets of real, non-zero, nonnegative and positive numbers \resp. $\N$ and $\N^*$ are the sets of nonnegative and positive integers \resp.
$l,m,n,p\in\N$ designate vectors and matrices dimensions. The subscript $k\in\N$ is the discrete time step and $i,j\in\N^*$ are vector and matrix component indices. 

\item $x_i$ is the $i$\ith component of the vector $\x$.  $a_{ij}$ is the $i$\ith row and $j$\ith column  element of $A\eqg\big[a_{ij}\big]=[\ab_j]\in\Rnm$ and
$\ab_j\in\Rn$ is its $j$\ith column vector  (if $n=0$ or $m=0$, $A$ is an empty matrix).

\item $\bff{0_n}\in\R^n$ and $0_{m\times n}\in\Rnm$ are zero vector and zero matrix resp. and $I_n\in\Rnn$ is the identity matrix.
\item $A^T$, $A^\dag$, $\rank(A)$, $\nul (A)$ and $\range (A)$ stand \resp. for the transpose, 
Moore-Penrose inverse, rank, kernel and range of the matrix $A$. If A is square, $\tr(A)$, $\abs{A}=\det(A)$ and $A^{-1}$, are its trace, determinant and inverse (if any) \resp.  
\item $\diag(x_i)_{i\in\{1,\cdots,k\}}$ is a diagonal matrix where  $x_1,\ldots,x_k$ are its diagonal elements.
%
%
 
%
\item A {Symmetric} matrix $M$ is {Positive Definite}, denoted by {SPD} or $M>0$ (\resp.   {Positive Semi-Definite} or non-negative definite, denoted by {SPSD} or $M\geq 0$) if and only if $\forall\x\in\Rn\text{--}\{\zero\}$, $\x^TMx>0$ (\resp.   $\x^TMx\geq~0$). This condition is met if and only if all its eigenvalues are real (because of its symmetry) and positive (\resp.   non-negative). The matrix inequality $M>N$ (\resp. $M\geq N$) means that  $M-N>0$ (\resp. $M-N\geq 0$).
\item 
 $\norm{\x}\eqg\norm{\x}_2\eqg\sqrt{\x^T\x}$ is the 2-norm of the vector $\x$; $\norm{A}_{2,1}\eqg\d\sum_j\norm{\ab_j}$ and $\norm{A}\eqg\norm{A}_2\eqg$ 
 $\sigma_{\max}(A)$. 
\item\label{unit_ball} 
$\BC^n_p\eqg\{\zb\in\R^n|\norm{\zb}_p\leq 1\}$ is a unit ball in $\R^n$ for the $p-$norm. 
$\BC^n_2$ and $\BC^n_\infty\eqg[-1,1]^n$ are the centred unit hypersphere and hypercube/box \resp.
%
\item\label{Minkowski} 
$\SC_1\oplus\SC_2\eqg\{\x\in\R^n|\x=\x_1+\x_2,\x_1\in\SC_1,\x_2\in\SC_2\}$ is the Minkowski sum of the sets $\SC_1,\SC_2\subset\R^n$ and 
%
$\mink_{i=1}^m\SC_i\eqg\SC_1\oplus\cdots\oplus\SC_m$.
\item\label{Ellipsoid} $\EC(\cb,P) \eqg \{x \in \R^n|\ (x-\cb)^T P^{-1}(x-\cb)\leq 1\}$ is an ellipsoid in $\R^n$, where $\cb\in \R^n$ is its center and $P\in\R^{n\times n}$ is a {\bf SPD} matrix that defines its shape, size and orientation in the $\R^n$ space. 
It can be also viewed as an affine transformation of matrix $M$ (where $M^TM=P$) of the unit Euclidean ball $\BC_2^n$: $\EC(\cb,M^TM)=~\{\x \in \R^n|\ \x=\cb+M\zb, \zb\in\BC_2^p\}$. If $M$ is not SPD but only {\bf SPSD}, then the ellipsoid is degenerate. It has an empty interior in the case where $p<n$.

%
%
\item\label{Hyperplane} $\HC(\db ,a)\eqg\{\x \in \R^n|\x^T\db =a\}$ is a hyperplane in $\Rn$ of normal vector $\db \in\Rn$ and whose signed distance from the origin is $\frac{{a}}{\norm{\db }}$.
Let also ${\GC}(\db ,a)\eqg\{\x: \x^T\db \leq a\}$ be one of the two halfspaces into which the hyperplane divides the $\R^n$ space and ${\GC}(-\db ,-a)$  
 is the other one.  Now let $\DC(\db ,a)\eqg{\GC}(\db,1+a)\cap{\GC}(-\db,1-a)$, \ie, $\DC(\db ,a)\eqg\{\x\in\R^n: \abs{\x^T\db-a}\leq 1\}$
which is the strip of $\R^n$,  of width ${2}{\norm{\db }}^{-1}$, that can also be seen as a degenerate unbounded ellipsoid or zonotope centred at $\HC(\db ,a)$.
%
$\PC(C,\db )=\bigcap_{i=1}^m\GC(\cb_i,d_i)$ is a polyhedron.
\item\label{Zonotope} $\ZC(\cb,L)\eqg \{\x \in \R^n|\ \x=\cb+L\zb, \zb\in\BC_\infty^m\}$$=\mink_{j=1}^q\{t_j\bm{l}_{j},\abs{t_j}\leq 1\}\oplus\{\cb\}$ is a zonotope of center $\cb$, obtained by affine transformation, of shape matrix $L\in\R^{n\times m}$, of the unit box $\BC_\infty^m$, 
where $m$ can be smaller, equal to or greater than $n$. A zonotope is also a convex polyhedron with centrally symmetric faces in all dimensions. 
Besides its vertex representation suitable for geometrical sum, there is a halfspace representation, suitable for intersection: $\ZC^\HC(D,\ab)=\bigcap_{i}\DC(\db_i,a_i)$.
%
\item\label{Support function} The support function of a set $\SC\subset\Rn$ is $\rho_\SC~:\Rn\rightarrow\R$, \linebreak
$\d\x\mapsto\rho_\SC(\x)\eqg\sup_{\ub\in\SC}\u^T\x$.  $\HC\big(\x,\rho_\SC(\x)\big)$ is the supporting hyperplane of $\SC$ and $\SC\subset \GC\big(\x,\rho_\SC(\x)\big)$. 
$\d\rho_{\EC(\cb,P)}(\x)=\cb^T\x+\sqrt{\x^TP\x}$ \cf \cite{Che:94}.
\end{enumerate}
}

\section{\uppercase{Problem formulation}}\label{sec_prob_form}
%
Consider the following linear discrete time 
system 
\mathl
\begin{subequations}\label{system0}
\begin{align}
&&\xka  &= \Akk \xkka+B\kkk \taub\kkk+R\kkk\nubkk, \label{state_eq0}\\
\text{where }\ &&\x_{0}&\in\EC(\xz,\sigz P_0)\eqd\EC_0\subset\Rn\text{ and }\nubk\in\BC_{\infty}^m, \label{init_state_noise_bound}
\end{align}
\end{subequations}
where $\xka\in\Rn$, $\taubk\in\R^{l}$ and $\nubk\in\Rm$  
are the unknown state vector to be estimated, a known and bounded control vector and an unobservable bounded process noise vector with unknown statistical characteristics, \resp., $\EC(\xz, \sigz P_0)\eqd\EC_0$ is a known ellipsoid (\cf $\S$\ref{sec_intro}.\ref{Ellipsoid}),  where $\xz\in\Rn$ is the initial estimate of $\xka$ at $k=0$,
$P_0\in\Rnn$ is a SPD matrix, $\sigz\in\R_+^*$ is a scaling positive scalar, the product $\sigz P_0$ is chosen as large as the confidence in $\xz$ is poor; $\BC_{\infty}^m$ is the unit ball for the $\infty-$norm in $\Rm$ (\cf $\S$\ref{sec_intro}.\ref{unit_ball}); $A_k\in\Rnn$ and  $B_k\in\Rnl$  are known state and input matrices, \resp. and $\Rk\in\Rnm$ is the generator matrix defining the shape of the zonotope bounding the unknown input vector:
$\etabk\in\ZC(\zero_n,R_k), \text{ where }\etabk\eqg R_k\nubk$. 
Now consider the output equation for the system \eqref{system0}: 
\begin{subequations}\label{bounds0}
\mathc
\begin{align}
&F_k^T\xka = \yk, \quad \yk\in\R^{p_k}\\
&\uyki\leq \yki\leq \byki,\ i\in\pscrk\eqg\{1,\ldots,p_k\}\subset\N,
\end{align}
\end{subequations}
where, the output matrix $F_k\eqg[\fbk{1}\ldots,\fbk{p_k}]\in\R^{n\times p_k}$ 
is time varying and so is the number of its columns, $p_k\in\N$, 
which can be zero sometimes. Indeed, the measurements are available in varying amounts, at not all but only some sporadic, not a priori known, time steps $k$. Three cases can be exhaustively enumerated: 
1) for some $i\in\dscrk\subset\pscrk$, both (finite and distinct) bounds are available: $\uyki<\byki$; 
2) for some $i\in\gscrk\eqg(\ugscrk\cup\bgscrk)\subset\pscrk$, only one bound, either  $\byki$ (if $i\in\bgscrk$) or  $\uyki$  (if $i\in\ugscrk$)  is available, in this case, the other (unavailable) bound is considered as $\mp\infty$.
%
3) for some other $i\in\hscrk\subset\pscrk$, the bounds are equal:  $\uyki=\byki$. The sets $\dscrk$, $\bgscrk$, $\ugscrk$ and $\hscrk$ form a partition for $\pscrk$: $\pscrk=\dscrk\cup\bgscrk\cup\ugscrk\cup\hscrk$. 
The measurement inequalities \eqref{bounds0} can be rewritten as
\begin{subequations}\label{bounds}
\mathc
\begin{align}
\fbki^T\xka&\leq\byki\text{ and } \uyki\rightarrow-\infty,&\text{if }i&\in\bgscrk, \label{boundb_HSpace0}\\
 -\fbki^T\xka&\leq-\uyki\text{ and } \byki\rightarrow+\infty,&\text{if } i&\in\ugscrk, \label{boundu_HSpace0}\\
\fbki^T\xka&=\byki,\text{ and } \uyki=\byki\ &\text{if } i&\in\hscrk,\label{bound_HyperP0}\\
\abs{\tfrac{1}{\gammaki}\fbki^T\xka-\yki}&\leq1,&\text{ otherwise } (i&\in\dscrk)\\
\text{where } \gammaki&\eqg\tfrac{\byki-\uyki}{2}\text{ and } \yki\eqg\tfrac{\byki+\uyki}{2\gammaki}&\label{bound_Strip0}
\end{align}
\end{subequations}
\begin{subequations}\label{output}
\begin{align}
\eqref{boundb_HSpace0}&\eq \xka\in\bGC\ki\eqg\GC(\fbki,\byki),&\forall i&\in\bgscrk, \label{boundb_HSpace}\\
\eqref{boundu_HSpace0}&\eq \xka\in\uGC\ki\eqg\GC(-\fbki,-\uyki),&\forall i&\in\ugscrk, \label{boundu_HSpace}\\
\eqref{bound_HyperP0}&\eq \xka\in\HC\ki\eqg\HC(\fbki,\byki),&\forall i&\in\hscrk,\label{bound_HyperP}\\
\eqref{bound_Strip0}&\eq\xka\in\DC\ki\eqg\DC\Big(\tfrac{1}{\gammaki}\fbki,\yki\Big),&\forall i&\in\dscrk,\label{bound_Strip}
\end{align}
where $\GC$, $\HC$ and $\DC$ are a halfspace, a hyperplane and a strip resp. (\cf $\S$\ref{sec_intro}.\ref{Hyperplane}).
\end{subequations}
\begin{assump}\label{assum}
From now on, we assume that
\begin{enumerate}
\item\label{assum_all_matrices_bounded} all known matrices and vectors intervening in \eqref{system0} and \eqref{bounds}, 
as well as 
 the SPD  $P_0$ and $\sigz\in\R_+^*$  
 are bounded;
\item\label{assum_all_matrices_nonzero} all the columns of all the matrices intervening in \eqref{system0} 
and those of $F_k$ are nonzero; 
\item\label{assum_uF_full_column_rank}  the matrix $F_k({\hscrk})\eqg[\fbki]_{i\in{\hscrk}}$, intervening in \eqref{bound_HyperP0} or \eqref{bound_HyperP}, has full column rank (in order to avoid contradictory constraints leading to an empty set);
\end{enumerate}
\end{assump}
\begin{aim}
We are intending here to design an estimator $\xk$ for the state  vector $\xka$  of the system \eqref{system0}-\eqref{bounds0}, such that, 
\begin{enumerate}
\item\label{requirement1}
a set (ellipsoid $\EC_k$ of center $\xk$) containing all possible values of the true state vector
$\xka$ is quantified, at each time step $k\in\N^*$ (standard requirement for a set-membership approach);
\item\label{requirement2} 
the state estimate vector $\xk$ is \emph{acceptable}, \ie, it belongs to all the sets defined in \eqref{bounds}.
%
\item\label{requirement3} 
under some conditions, the estimator $\xk$ is ISS, (Input-to-State Stable, \cf  Theorem \ref{lemm_ISS}). 
This is one of the distinguishing features of the algorithm designed here.
\end{enumerate}
\end{aim}
The other distinguishing feature is that, unlike the other set-membership techniques, such as those using exclusively intervals, zonotopes or polytopes, the one detailed here delivers an optimal (\wrt some chosen criterions) set, without any conservatism. 
Since the only measured information about the true state vector $\xka$ consists in its belonging to the sets defined in \eqref{bounds},  there is no better estimate than the one that belongs to these sets. But such an estimator is not unique and is not necessarily stable so the most suitable one will be
chosen among the set of all possible estimators by optimizing a given criterion. 

Let $\EC_{k}\eqg\EC(\xk,\sigk P_{k})$ be the ellipsoid containing all possible values of the true state vector
$\xka$. 
Note that the singular values 
 of the shape matrix $\sigk P_{k}$ 
correspond to the semi-lengths of its axes, whose directions are
defined by the associated--orthogonal since $P_{k}$ is symmetric--eigenvectors.
In what follows, we have to determine the progression law for the
ellipsoid $\EC_{k}$ (and thence for the state
estimate vector $\xk$) such that the aims \bff{i}.--\bff{iii}. are
fulfilled. 
\section{ Time update (prediction stage)\ }\label{sec_time_update}
%
Let $\EC\kk\eqg\EC(\xkk,\sigkkk\Pkk)$ be the ellipsoid including the
\guil{reachable set} 
of every possible value of $\xkka\in\EC\kkk$ that
evolves according to the plant dynamics eq. \eqref{state_eq0}, subject to \eqref{init_state_noise_bound}. The following theorem gives the parametrized family of ellipsoids $\EC\kk$ that contains the sum of the ellipsoid resulting of the endomorphism of matrix $A\kkk$ applied to $\EC\kkk$ on one hand and the zonotope $\ZC(\zero_n,{R}\kkk)$, on the other.
\begin{theo}\label{theo_time_update}
If $\xkka\in\EC\kkk\eqg\EC(\xkkk,\sigkkk\Pkkk)$ and $\xka$ obeys to \eqref{system0}, then $\forall\mub\eqg(\mu_1,\ldots,\mu_m)^T\in\big(\R_+^{*}\big)^m$,%
$$\xka\in\EC(\xkk,\sigkkk\Pkk)\eqd\EC\kk\eqg\EC\kkj{m}\supseteq\EC\kkj{m-1}\supseteq\ldots\supseteq\EC\kkj{0}$$
where  
$\EC\kkj{i}\eqg \EC(\xkk,\sigkkk\Pkki)$ and 
\mathc
\begin{subequations}\label{predic_eq}
 \begin{align}
\xkk&\eqg\Akk\xkkk+B_{k-1}\taubkk,\label{x_pred}\\
\Pkk&\eqg P_{{k/k-1}_m},\label{P_pred}\\
 P_{{k/k-1}_{0}}&\eqg\Akk\Pkkk \Akk^{T};\label{P_pred0}
\end{align}
and, $\forall i\in\{1,\ldots,m\}$,
\mathc
\begin{align}
\Pkki&\eqg(1+\mu_i)P_{k/k-1_{i-1}}+\frac{1+\mu_i}{\mu_i\sigkkk}\rkki\rkki^{T},\label{P_pred1}
\end{align}
\end{subequations}
$\rki$ (the $i$\ith column of $\Rk$) 
being the generator vector of the zonotope
containing all possible values of the process noise $\etabk\eqg R_k\nubk$.
\end{theo}
{\debdemo 
Conforming to \eqref{state_eq0}, the set containing every possible value of $\xka$ can be schematized by 
\mathc
\begin{align}
\big(A\kkk\EC\kkk\oplus\big\{B_{k-1}\taubkk\big\}\big)\oplus\ZC(\zero_n,R_k).
\end{align}
First, $A\kkk\EC\kkk=\EC(\Akk\xkkk,\sigkkk\Akk P_{k-1}\Akk^{T})$ is the  image of the ellipsoid $\EC_{k-1}$ by the endomorphism of matrix $\Akk$ and $\EC_{{k/k-1}_{0}}$ 
 is its translation by the known vector $B_{k-1}\taubkk$.
Secondly, $\EC\kk$ is the outer-bounding ellipsoid of the Minkowski sum (\cf $\mathsection$.\ref{sec_intro}.\ref{Minkowski})
of $\EC_{{k/k-1}_{0}}$ and the zonotope $\ZC(\zero_n,\Rkk)$:
$\EC\kk\supset\big(\EC_{{k/k-1}_{0}}\oplus\ZC(\zero_n,\Rkk)\big)$.
%
Thirdly, the zonotope $\ZC(\zero_n,\Rk)$, where $\Rk=\big[\r_{k_1},\cdots,\r_{k_m}\big]$, can be represented as the sum of $m$ degenerate ellipsoids \cite{kur:14}: $$\d\ZC(\zero_n,\Rk)=\mink_{i=1}^m\EC(\zero_n,\r_{k_i}\r_{k_i}^T).$$
Now, the Minkowski sum of two ellipsoids $\EC(\cb_{1},P_{1})$  and $\EC(\cb_{2},P_{2})$ is not an ellipsoid, in general, yet can be bounded by a parametrized ellipsoid \cite{Mak:96b}: 
\mathc
\begin{align}
\EC(\cb,P(\mu))&\supset\EC(\cb_{1},P_{1})\oplus\EC(\cb_{2},P_{2}), \ \forall\mu\in\R_+^*, \\
\text{ where }\cb&=\cb_{1}+\cb_{2}\text{ and }P(\mu)=(1+\mu)P_{1}+(1+\sfrac{1}{\mu})P_{2}.
\end{align}
 Applying this result sequentially to 
\mathc
\begin{align}
\d\EC\kk\supset\EC_{{k/k-1}_{0}}\oplus\left(\mink_{i=1}^m\EC(\zero_n,\rkki\rkki^T)\right),
\end{align} 
eventuates in \eqref{predic_eq}. 
\findemo 
}

The parameters $\mu_i\eqg \mu_{k_i}$ are positive scalars, chosen in such a way as to minimize the size of $\EC\kk$. The most telling measures of the size of an ellipsoid are surely the volume and the sum of the squared semi-axes lengths. Since the eigenvalues of $\sigkkk\Pkk$ are the squared semi-axes lengths of $\EC\kk$, the former is proportional to their product, \ie, to $\sigkkk^n\det(\Pkk)$ and the latter is equal to $\sigkkk\tr(\Pkk)$.

\begin{theo}\label{theo_mu-vol-tr} Let $\EC\kkp$ defined in Thm \ref{theo_time_update}. 
\nopagebreak
\begin{enumerate}
\item 
 If $P\kkpj{0}$ is SPD, then $\EC\kkp$ has the minimum volume if $\mub=\mub_{k}^{\text{vol}}$, where, $\forall  i\in\{1,\cdots,m\}$,
\mathc
\begin{subequations}\label{mu-vol}
\begin{align}
\mu_{k_{i}}^{\text{vol}}&\eqg\tfrac{1}{2n}\sqrt{(n-1)^{2}{h}_{k_{i}}^{2}+4n{h}_{k_{i}}}-\tfrac{n-1}{2n}{h}_{k_{i}},\\
\intertext{ where }
{h}_{k_{i}}&\eqg\sigk^{-1}\rkit P\kkpj{i-1}^{-1}\rki.\label{h_def} 
\end{align}
\end{subequations}
\item  $\EC\kkp$ has the minimum  sum of the squared axes lengths, if $\mub=\mub_{k}^{\text{tr}}$, where
\mathc
\begin{align}\label{mu-tr}
\mu_{k_{i}}^{\text{tr}}&\eqg\sqrt{\tfrac{\rkit \rki}{\sigk\tr(P\kkpj{i-1})}}, \forall  i\in\{1,\cdots,m\}.
\end{align}
%
and the recursive formula \eqref{P_pred}-\eqref{P_pred1} becomes: 
\begin{subequations}\label{predic_eq_glo}
\mathc
\begin{align}
P\kkp&=\big(1+\tfrac{\bar{\mu}_{k}}{\bar{\mu}_{k_0}}\big)\big(P\kkpj{0}+\tfrac{\bar{\mu}_{k_0}}{\sigk}\bar{R}_{k}\big),\label{P_pred_glo}\\
\intertext{\text{where } }
 \bar{\mu}_{k_0}&\eqg \sqrt{\sigk\tr(P\kkpj{0})},\text{ where $P\kkpj{0}$ given in \eqref{P_pred0},}\label{mu0_def}\\
\bar{\mu}_{k}&\eqg\sum_{i=1}^m {\norm{\rki}}= \norm{R_k}_{2,1},  
\label{mu_def}\\
\bar{R}_{k}&\eqg\sum_{i=1}^m\norm{\rki}^{-1}\rki\rki^T.\label{bR_def}
\end{align}
\end{subequations}
\end{enumerate}
\end{theo}
{{\debdemo 
\begin{enumerate}
\item If  $P_{k/k-1_0}$ is SPD then $P_{k/k-1_i}$ is so, $\forall i\in\{1,\ldots,m\}$.  
The volume of an ellipsoid being proportional to the determinant of its shape matrix and $\sigkkk$ being considered as constant at time step $k$,
$\d\mu_{k_{i}}^{\text{vol}}=\arg\min_{\mu_i}\det(P\kkpj{i})$, \eqref{mu-vol} can be deduced from \cite{Che:94} p. 84.
\item The sum of the squared semi-axes lengths of an ellipsoid being the trace of its shape matrix,
$\d\mu_{k_{i}}^{\text{tr}}=\arg\min_{\mu_i}\tr(P\kkpj{i})$;
\eqref{mu-tr} can be derived from \cite{Mak:96b}.  As for \eqref{predic_eq_glo}, it is a direct consequence of the result \cite{Dur:01} saying that the minimum trace ellipsoid containing the Minkowski sum of $m$ ellipsoids is : 
\mathc
\begin{align}
\EC(\cb,P)&\eqg\oplus_{i=1}^m\EC(\cb_i,P_i),\\ 
\intertext{where }
\cb\eqg \sum_{i=1}^m \cb_i \text{ and }
P&\eqg \Big(\sum_{i=1}^m\sqrt{\tr(P_i)}\Big) \Big(\sum_{i=1}^m\big(\sqrt{\tr(P_i)}\big)^{-1}{P_i}\Big).\label{pf_theo_mu-vol-tr_1}
\end{align}
Then, after noticing that 
\mathc
\begin{align}
\tr(\rkki\rkkit)=\rkkit\rkki=\norm{\rkki}^2\\
\intertext{and that}
\sum_{i=1}^m\frac{\rki\rki^{T}}{\sqrt{\tr(\rkki\rkkit)}}=R_k\diag(\bar{\mu}_{k_i})_{i\in\{1,\cdots,m\}}^{-1}R_k^T, 
\end{align}
\eqref{pf_theo_mu-vol-tr_1} applied to $\d\EC\kkj{0}\oplus\big(\oplus_{i=1}^m\EC(\zero_n,\rkki\rkkit\big)$, leads clearly to \eqref{predic_eq_glo}.
It is also stated in \cite{Dur:01} that such an ellipsoid is the same that the one obtained sequentially in \eqref{mu-tr}.\findemo 
\end{enumerate}
}
}
\begin{rema}
It is worth noting that the volume minimisation problem \linebreak $\d\arg\min_{\mu_i}\det(\Pkki)$ has an explicit solution here.  If the unknown input vector was bounded by an ellipsoid, as was the case in \cite{Mak:96b,Dur:01,Bec:08,She:18}, rather than by an interval-like set, such as a zonotope, $\d\mu_{k_{i}}^{\text{vol}}$ would be the unique positive solution of an $n-$order polynomial to be solved at each time step $k$. 
\end{rema}
\begin{rema}
Because of the equality constraints introduced by  the measurements $i\in\hscrk$, the matrix $\Pk$ looses rank during the correction stage. Hence, the $\EC\kk$'s volume minimization, \eqref{mu-vol}, should be avoided in favor of \eqref{mu-tr} because of the inversion of $\Pkki$ involved in the former, in \eqref{h_def}. 
\end{rema}
\begin{rema}
When minimizing the sum of squared axes lengths of $\EC\kk$, it is not necessary to compute the $m$ intermediate values of $\Pkki$, given by the recursive formula \eqref{P_pred}-\eqref{P_pred1}. $\Pkk$ can be computed directly using \eqref{predic_eq_glo} instead. Note also that the matrix $M_k$ is invertible thanks to the assumption \ref{assum_all_matrices_nonzero}
\end{rema}
\begin{rema}
It is possible to minimize the weighted sum of the squared axes lengths of  $\EC\kk$: $\tr(C\Pkk C^T)$, for any $C\in\R^{n_C\times n}$, $n_C\in\N^*$. In this case, the optimal value for $\mub$ would be (\cf \cite{Che:99})
\mathc
\begin{align}\label{mu-tr-C}
\mu_{k_{i}}^{\text{tr}}&\eqg\sqrt{\frac{\rkkit C^TC\rkki}{\sigkkk\tr(C P_{k/k-1_{i-1}}C^T)}},\  i\in\{1,\cdots,m\},\\
\bar{\mu}_{k}&\eqg\norm{CR_kR_k^TC^T}_{2,1}= \sum_j^m\norm{C\rki}, \\ 
{M}_{k}&\eqg\diag(\norm{C\rki})_{i\in\{1,\cdots,m\}}.
\end{align}
\end{rema}

Given the ellipsoid at the previous time step $\EC\kkk$, Thm \ref{theo_time_update} provides the ellipsoid $\EC\kk$ whose center is given by \eqref{x_pred} and whose shape matrix is given, up to the factor $\sigkkk$, by the recursive formula \eqref{P_pred}-\eqref{P_pred1} which depends on $\mub$; Thm \ref{theo_mu-vol-tr} offers the optimal values for this parameter according to two criterions, the choice of which is let to the user, in the absence of equality constraints. Otherwise, 
the shape matrix is calculated directly 
by \eqref{predic_eq_glo}. 
%
%
\section{ Measurement update (correction)\ }\label{sec_measurement_update}
%
The dynamic state evolution equation \eqref{system0} allowed to compute the predicted ellipsoid $\EC\kk$ which contains all possible values of the state vector $\xka$ taking into account all the measurements up to time step $k-1$ if any. Now, let us recall the other sets containing $\xka$, obtained from the measurements:
\mathc%
\begin{align}
\eqref{bounds}  &\eq\xka\in\bigcap_{i\in\gscrk}\GC\ki\cap\bigcap_{i\in\dscrk}\DC\ki\cap\bigcap_{i\in\hscrk}\HC\ki,\text{ if } p_k\neq 0.\label{set_XC_P_eq}
\end{align}
It is interesting to note that the intersection of half-spaces can be considered as a possibly unbounded polyhedron and that the intersection of strips is a zonotope:  
\begin{align}
\bigcap_{i\in\gscrk}\GC\ki&\eqd\PC_k\eqg\PC\big([\fbki]_{i\in\gscrk},[\yki]_{i\in\gscrk}\big),\\
\bigcap_{i\in\dscrk}\DC\ki&\eqd\ZC_k\eqg\ZC^{\HC}\big([\fbki]_{i\in\dscrk},[\yki]_{i\in\gscrk}\big).
\end{align}
The correction stage consists in performing the intersection between $\EC\kk$ and the set \eqref{set_XC_P_eq}, allowing to find $\EC_k\supset\SC_k$ in light of the current measurements, where
\mathc
\begin{align}
\SC_k&\eqg\Big(\big(\EC\kk\cap\bigcap_{i\in\gscrk}\GC\ki\big)\cap\bigcap_{i\in\dscrk}\DC\ki\Big)\cap\bigcap_{i\in\hscrk}\HC\ki\\
&=\Big(\big(\EC\kk\cap\PC_k\big)\cap\ZC_k\Big)\cap\bigcap_{i\in\hscrk}\HC\ki.
\label{SC_set} 
\end{align}
It will be shown that this intersection is the one between $\EC\kk$ and the possibly degenerate (if $\hscrk\neq\emptyset$) zonotope.
It does not result in an ellipsoid in general and has to be circumscribed by such a set, which is the subject of the upcoming paragraphs.
%
We shall begin by working on the intersection $\d\EC\kk\cap\GC\ki$ in $\S$\ref{subsec_poly}.
Secondly, we'll be dealing with the intersection between an ellipsoid and a  strip in order to carry out the set obtained in $\S$\ref{subsec_poly} and intersecting it with $\d\bigcap_{i\in\dscrk}\DC\ki$;  
$\S$\ref{subsec_strip} provides the optimal ellipsoid overbounding this intersection.
Thirdly, the intersection of an ellipsoid with a hyperplane will be presented in $\S$\ref{subsec_hyp}, in order to handle the intersection of the previously obtained ellipsoid with $\d\bigcap_{i\in\hscrk}\HC\ki$.
Finally, all these results will be compiled in a unique state estimation algorithm in $\S$\ref{sec_algo}.
\subsection{ Intersection of an ellipsoid with a halfspace}\label{subsec_poly}
The intersection between the ellipsoid $\EC\kk$ obtained in $\mathsection$~\ref{sec_time_update} and the polyhedron $\PC_k$ 
can be reformulated as the intersection of $\EC\kk$ and a series of strips $\DC\ki$.
To grasp this idea, take any closed convex set  $\SC$ and a hyperplane $\HC$ intersecting it. The intersection of $\SC$ with a halfspace $\GC$ delimited by $\HC$ is nothing else that its intersection with the strip formed between $\HC$ and a support hyperplane of $\SC$, parallel to $\HC$ and contained in $\GC$. Now, if $\HC$ doesn't intersect $\SC$, the latter is either a subset of $\GC$ or lies outside of it, and if $\HC$ is tangent to $\SC$ (being its support hyperplane), then $\SC$ is either again a subset of $\GC$  or it has only one point in common with it. 
In the case where $\SC$ is an ellipsoid and the intersecting halfspace corresponds to the constraint \eqref{boundb_HSpace}/\eqref{boundb_HSpace0}, the theorem below provides the parameters of the intersecting strip. 
To obtain the intersection of an ellipsoid with the halfspace given by the constraint \eqref{boundu_HSpace}/\eqref{boundu_HSpace0}, it suffices to replace $\fb$ by $-\fb$ and  $\by$ by $\uy$:
\begin{theo}[ellipsoid-halfspace intersec.]\label{theo_ellips_HSpace_inter} Let 
$\cb\in~\Rn$, $\fb\in\Rn\text{--}\{\zero_n\}$, $P\in\Rnn$ SPSD, $\varsigma\in\R_+^*$ and $\by\in\R$.
\mathc
\begin{subequations}
\begin{align}
\intertext{{If } $\by<-\urho$,\hfill{\normalfont{(case 1)}}}
&&\hspace{-5mm}\EC(\cb,\varsigma P)\cap\GC(\fb,\by)&=\emptyset;\label{cas1a}\\
\intertext{else if $\by\geq\brho$,\hfill{\normalfont{(case 2)}}}
&&\hspace{-5mm}\EC(\cb,\varsigma P)\cap\GC(\fb,\by)&=\EC(\cb,\varsigma P);\label{cas1b}\\
\intertext{else if 
$\by=-\urho$,\hfill{\normalfont{(case 3)}}}
&&\hspace{-5mm}\EC(\cb,\varsigma P)\cap\GC(\fb,\by)&=\EC(\cb,\varsigma P)\cap\HC(\fb,-\urho)
=\{\cb-\varsigma^{\frac{1}{2}}(\fb^TP\fb)^{-\frac{1}{2}} P\fb\};\hspace{-5mm}\label{cas3_HSpace}\\
\intertext{{else  (}$-\urho<\by<\brho$), \hfill\normalfont{(case 4)}}
&&\hspace{-5mm}\EC(\cb,\varsigma P)\cap\GC(\fb,\by)&=\EC(\cb,\varsigma P)\cap\DC(\tfrac{1}{\gamma}\fb,{y}),\label{case4_Hspace}\\
\intertext{where}
&&\gamma&\eqg \tfrac{1}{2}(\by+\urho)
\text{ and }{y}\eqg \tfrac{1}{2\gamma}(\by-\urho)\label{gamma_y_HSpace_def}\\ 
&&\urho&\eqg\rho_{\EC(\cb,\varsigma P)}(-\fb)=-\cb^T\fb+\sqrt{\varsigma\fb^TP\fb}\text{ (\cf $\S$\ref{sec_intro} \ref{Support function})}\hspace{-5mm}\label{urho_def}\\
&&\brho&\eqg\rho_{\EC(\cb,\varsigma P)}(\fb)=\cb^T\fb+\sqrt{\varsigma\fb^TP\fb}. \label{brho_def}
\end{align}
\end{subequations}
\end{theo}
{\debdemo 
\nopagebreak
\begin{defi}
The signed distance from a set $\SC\subset\Rn$ to a vector $\x\in\Rn$ is $\d\zeta(\SC,\x)\eqg\max_{\norm{\ub}=1}\ub^T\x-\rho_\SC(\ub)$.
\end{defi}
\begin{propo}[\cite{Kur:06}]\label{ellips_Hplane_dist}
The  signed distance from an ellipsoid  to a hyperplane is given by:
\begin{align}
\zeta\big(\EC(\cb,P),\HC(\db,a)\big)\eqg\norm{\db}^{-1}\Big(\abs{a-\cb^T\db}-\sqrt{\db^TP\db}\Big).
\end{align}
\end{propo}
Let $\EC\eqg\EC(\cb,\varsigma P)$. The signed distance from $\EC$ to $\HC(\fb,\by)$ is 
\begin{align}
\zeta\eqg\norm{\fb}^{-1}\left(\abs{\by-\cb^T\fb}-\sqrt{\varsigma\fb^TP\fb}\right).
\end{align}
\begin{itemize}
\item $\zeta\geq 0$ means that $\HC(\fb,\by)$ does not intersect $\EC$ in more than one point: 
\begin{enumerate}
\item if $\cb^T\fb>\by$, then $\EC\subset~\GC(-\fb,-\by)$ and $\EC\cap\GC(\fb,\by)=\emptyset\eq$ \eqref{cas1a}; 
\item if $\cb^T\fb\leq \by$, then $\EC\subset\GC(\fb,\by)\Rightarrow\EC\cap\GC(\fb,\by)=\EC\eq$ \eqref{cas1b};
\item if $\cb^T\fb-\by=\sqrt{\varsigma\fb^TP\fb}$, then $\HC(\fb,\by)$ is tangent to $\EC$
 and   \linebreak
 $\EC\cap\GC(\fb,\by)=\EC\cap\HC(\fb,\by)=\{\check{\cb}\}$, where $\check{\cb}$ is calculated using \eqref{eq_x},  letting $\delta\leftarrow \by-\cb^T\fb$, with
 $\delta^2={\varsigma\fb^TP\fb}$, $\fb\leftarrow\fb$, $P\leftarrow\varsigma^{-1} P$.
\end{enumerate}
\item If $\zeta\leq0$, then $\HC(\fb,\by)$ intersects $\EC$ and
$\HC(-\fb,\rho(-\fb))$ is the ellipsoid's supporting hyperplane of normal vector $-\fb$ which is contained in $\GC(\fb,\by)$.
Indeed, 
\begin{align}
\x\in\HC(-\fb,\rho(-\fb))&\eq\x^T\fb-\cb^T\fb=-\sqrt{\varsigma\fb^TP\fb}\leq \by-\cb^T\fb\nn\\
&\Rightarrow \x^T\fb\leq \by\eq\x\in\GC(\fb,\by).\\
%
\intertext{Thence, $\GC\big({-\fb},\rho(-\fb)\big)$ is its supporting halfspace and $\EC\subset\GC\big(-\fb,\rho(-\fb)\big)$.
Therefore, }
\EC\cap\GC(\fb,\by)&=\Big(\GC\big({-\fb},\rho(-\fb)\big)\cap\GC(\fb,\by)\Big)\cap\EC\\
\intertext{$\abs{\by-\cb^T\fb}<\sqrt{\varsigma\fb^TP\fb}$ means that $0<\by+\rho(-\fb)<2\sqrt{\varsigma\fb^TP\fb}$, 
entailing, on one hand, }
\GC(\fb,\by)&=\big\{\x\big|\x^T\fb\leq \by\big\}=\bigg\{\x\bigg|\frac{2}{\by+\rho(-\fb)}\x^T\fb\leq\frac{2\by}{\by+\rho(-\fb)}\bigg\}\nn\\ &=\GC(\gamma^{-1}\fb,{y}+1)\\
\text{and }
\GC(-\fb,\rho(-\fb))&=\bigg\{\x\bigg|-\frac{2}{\by+\rho(-\fb)}\x^T\fb\leq\frac{2\rho(-\fb)}{\by+\rho(-\fb)}\bigg\}\nn\\
&=\GC(-\gamma^{-1}\fb,-{y}+1),\\
\intertext{on the other hand. Finally, the proof \eqref{case4_Hspace}--\eqref{gamma_y_HSpace_def} is achieved thusly:}
\EC\cap\GC(\fb,\by)&=\EC\cap\Big(\GC\big(\gamma^{-1}\fb,{y}+1\big)\cap\GC\big(-\gamma^{-1}\fb,-{y}+1\big)\Big)\\
&=\EC\cap\DC\big(\gamma^{-1}\fb,{y}\big).\tag*{\ding{113}}
\end{align}
\end{itemize}
The figure \ref{ell_hs_inters_fig} illustrates the above theorem. It shows how the intersection between a (blue/big) ellipsoid  and a halfspace (all the colored area) is the same as the intersection of this ellipsoid with a strip (dark colored area). The (red/small) ellipsoid overbounding this intersection will be calculated in the next paragraph.
\begin{figure}[htbp]
\begin{center}
\includegraphics[width=0.8\textwidth]{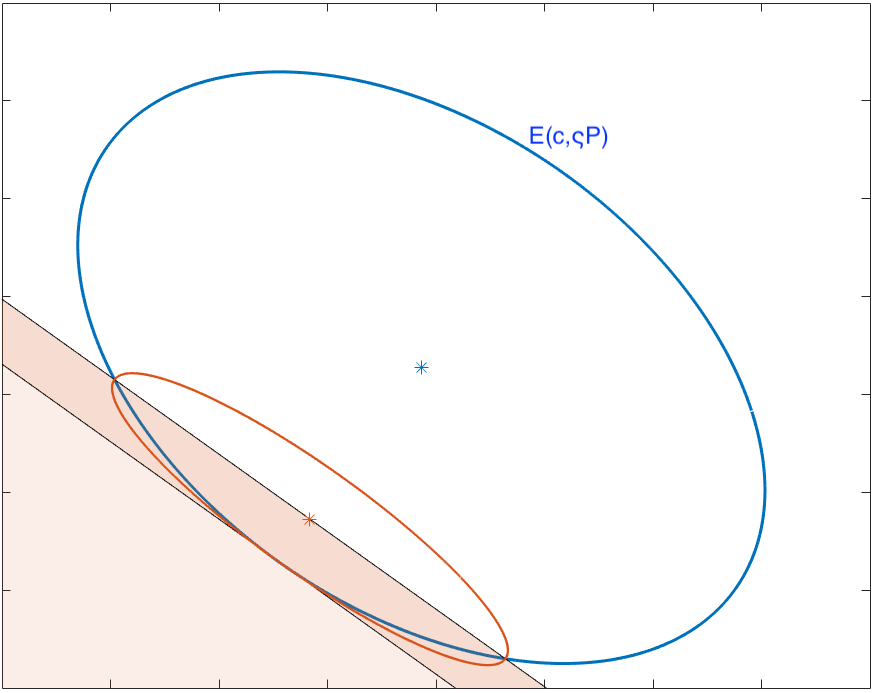}
\caption{Intersection of an ellipsoid with a halfspace ($n=2$)}
\label{ell_hs_inters_fig}
\end{center}
\end{figure}

}
\subsection{ Ellipsoid bounding the intersection of an ellipsoid with a strip} \label{subsec_strip} 
In the previous paragraph, we showed that the incorporation of the measurements $i\in\gscrk$ result, as for those $i\in\dscrk$, from the intersection of the predicted ellipsoid with a zonotope, formulated as an intersection of several strips. We need now to overbound this intersection by an ellipsoid. 
To begin with, the theorem below presents a family of parametrized ellipsoids that contain an ellipsoidal layer, coming out of the intersection of $\EC(\cb,\varsigma P)$ with the strip 
$\DC(\fb,y)$, which can be considered--interestingly enough--as an ellipsoid unbounded in all directions orthogonal to $\fb$.
\begin{theo}[ellips./strip inters.]\label{theo_ellips_strip_inter}
Let $\cb\in\Rn$,  $\varsigma\in\R_+^*$, $y\in\R$, $P\in\Rnn$ SPSD and $\fb\in\Rn\text{--}\{\zero_n\}$, \\
%
 \mathc
\begin{subequations}\label{lemm_ellips_strip_inter_eq}
\begin{align}
\intertext{if $y-1>\brho$ or $y+1<-\urho$,\hfill\normalfont{(case 1)}}
\hspace{-6pt}\DC(\fb,y)\cap\EC(\cb ,\varsigma P)&=\emptyset;\\
\intertext{else if $y+1\geq\brho$ and $y-1\leq -\urho$,\hfill\normalfont{(case 2)}}
\hspace{-6pt}\DC(\fb,y)\cap\EC(\cb ,\varsigma P)&=\EC(\cb ,\varsigma P);\\
\intertext{else if $y=-\urho-1$,\hfill\normalfont{(case 3.a)}}
\hspace{-6pt}\DC(\fb,y)\cap\EC(\cb ,\varsigma P)&=\HC(\fb,-\urho)\cap\EC(\cb,\varsigma P)=\{\cb-\varsigma^{\frac{1}{2}}(\fb^TP\fb)^{-\frac{1}{2}}P\fb\};\!\!\!\!\!\\
\intertext{else if $y=\brho+1$,\hfill\normalfont{(case 3.b)}}
\hspace{-6pt}\DC(\fb,y)\cap\EC(\cb ,\varsigma P)&=\HC(\fb,\brho)\cap\EC(\cb,\varsigma P)=\{\cb+\varsigma^{\frac{1}{2}}(\fb^TP\fb)^{-\frac{1}{2}}P\fb\};
\end{align}
\end{subequations}
\mathc
\begin{subequations}\label{Ellips_Strip_Inter}
\begin{align}
\intertext{else $(-\urho< y+1<\brho$ or $-\urho< y-1<\brho), \ \forall\beta\in]0,1[$,\hfill\normalfont{(case 4)}}
\hspace{-5mm}\DC(\fb,y)\cap\EC(\cb ,\varsigma P)=
\DC(\breve{\fb},\breve{y})\cap\EC(\cb ,\varsigma P)
\subset\EC(\breve{\cb }(\beta),\breve{\varsigma}(\beta)\breve{P}(\beta))\eqd\EC(\beta), \label{ellips_strip_inter}
\end{align}
%
\begin{align}
\intertext{where }
\breve{\fb}&\eqg\tfrac{1}{\gamma}\fb \text{ and } \breve{y}\eqg\tfrac{1}{\gamma}(\fb^T\cb+\delta), \label{breve_f_y_def}\\
\breve{P}(\beta)& \eqg  P- {\alpha\beta} P\fb\fb^T P,\quad      \label{P_ellips_strip_inter}\\
\breve{\cb}(\beta)   & \eqg  \cb +{\alpha\beta}\delta P\fb,  \label{c_ellips_strip_inter}\\
\breve{\varsigma}(\beta) & \eqg  \varsigma+{\alpha\beta}\left(\gamma^2({1-\beta})^{-1}-\delta ^2\right),\label{sig_ellips_strip_inter}\\ 
\alpha&\eqg \big(\fb^TP\fb\big)^{-1}, \label{alpha_def}\\
\delta&\eqg \tfrac{1}{2}(\by+\uy)-\fb^T\cb=\tfrac{1}{2}(\by+\uy-\brho+\urho),\label{delta_def}\\
    \gamma&\eqg \tfrac{1}{2}(\by-\uy),\label{gamma_def}\\ 
    \by&\eqg\min(y+1,\brho) \text{ and } \uy\eqg\max(y-1,-\urho) 
 \end{align}
  \end{subequations}
and  $\urho$ and $\brho$ are defined in \eqref{urho_def}--\eqref{brho_def}.
\end{theo}
{\debdemo 
Let $\EC\eqg\EC(\cb ,\varsigma P)$.
The signed distance from the ellipsoid $\EC$ to each of the two hyperplanes $\HC(\fb,y\mp1)$, bounding the strip $\DC(\fb,y)$, is 
\begin{align}
\zeta\eqg\norm{\fb}^{-1}\big(\big| y\mp 1-\fb^T\cb\big|-\sqrt{\varsigma\fb^TP\fb}\big).
\end{align}
When $\zeta>0$, the ellipsoid doesn't intersect any of both hyperplanes meaning either that it is situated between them \ie, contained in the strip (case 2) or that the ellipsoid is located outside the strip, in which case (case 1), the intersection is empty. 
In the case 3, the interior of the ellipsoid is outside the strip touching it in only one point and the case 3 of Thm \ref{theo_ellips_HSpace_inter} is then applicable: $\DC(\fb,y)\cap\EC=\GC(\fb,y+1)\cap\EC$ (case 3.a) and $\DC(\fb,y)\cap\EC=\GC(-\fb,-y+1)\cap\EC$ (case 3.b).
In the case 4,  where  $\zeta\leq~0$, the intersection is not empty. It is then possible to introduce the following lemma, based on the results of \cite{Fog:82} and \cite{Tan:97}:
\begin{lemm}
$\forall y\in\R$, $\bm{c}\in\Rn$, $\fb\in\R^{n}$, $\sigma\in~\R_+^*$ and SPD $P\in\Rnn$, if $\DC(\fb,y)\cap\EC(\bm{c},\sigma P)\neq\emptyset$, then 
\begin{align*}
\forall\omega\in\R_+^*,&\quad\EC(\tilde{\bm{c}}(\omega),\tilde{\sigma}(\omega)\tilde{P}(\omega))\supset\DC(\fb,y)\cap\EC(\bm{c},\sigma P),\\
\intertext{where }   \tilde{P}(\omega)& \eqg  P-\omega (\omega  \alpha + {1})^{-1} P {\fb}{\fb}^TP,\quad \\    
   \tilde{\bm{c}}(\omega)   & \eqg  \bm{c}+\omega (\omega  \alpha + {1})^{-1}\delta P {\fb}= \bm{c}+\omega  \tilde{P}(\omega) {\fb}^{-1}\delta,\quad   \\       
   \tilde{\sigma}(\omega) & \eqg  \sigma+\omega(1-\alpha(\omega+ \alpha)^{-1}\delta^2), \nn\\
   \delta&\eqg y-{\fb}^T\bm{c} \text{ and  $\alpha$ is given in \eqref{alpha_def}}.
   \end{align*}
\end{lemm}
This lemma is also the mono-output case of the \guil{observation update} part of 
Thm 1, \cite{Bec:08}. 
\eqref{P_ellips_strip_inter}, \eqref{c_ellips_strip_inter} and \eqref{sig_ellips_strip_inter} are obtained by setting \linebreak $\omega\eqg{\alpha\beta}({1-\beta})^{-1}$, thus $\beta={\omega}({\alpha+\omega})^{-1}$. 
%
%
But before applying the lemma above, it is suitable to reduce the strip $\DC(\fb,y)$ in case where one of the two hyperplanes does not intersect the ellipsoid $\EC$, \ie, when either $y+1>\brho$ or $y-1<-\urho$, by translating the aforementioned hyperplane so that it becomes tangent to the ellipsoid, as proposed in \cite{Bel:90}. The new strip so obtained is $\DC(\gamma^{-1}\fb,\breve{y})$, where $\gamma$ and $\breve{y}\eqg y$ are given in \eqref{gamma_y_HSpace_def} and obtained by applying (case 4) of Thm \ref{theo_ellips_HSpace_inter} to $\EC\cap\GC(-\fb,-y+1)$  and to $\EC\cap \GC(\fb,y+1)$.
 \findemo }

\subsection{ Optimal value of the  parameter $\beta$}\label{sec_beta_opt}
\begin{figure}[htbp]
\begin{center}
\includegraphics[width=0.8\textwidth]{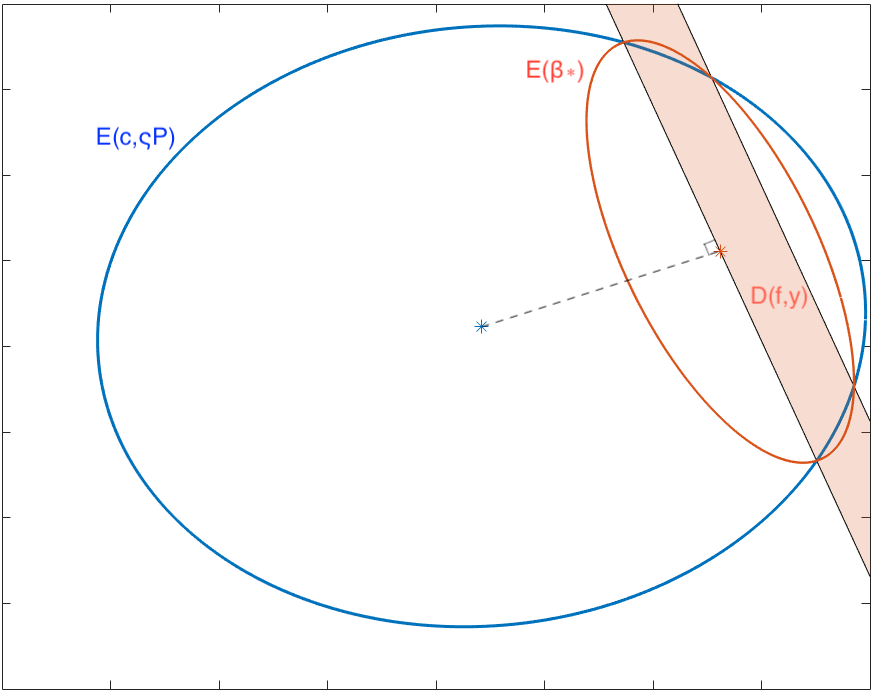}
\caption{Intersection of an ellipsoid with a strip ($n=2$)}
\label{ell_strip_inters_fig}
\end{center}
\end{figure}

\indent Now, the optimal value of the weighting parameter $\beta$ with respect to a
judiciously chosen criterion is derived. 
In their well-known paper  \cite{Fog:82}, Fogel and Huang give two optimal values of $\omega\eqg\dfrac{\alpha\beta}{1-\beta}$: the first minimizing the determinant of $\breve{\varsigma}\breve{P}$ and the second, its trace, 
thus optimizing  the volume and the sum,  \resp., of the squared semi-axes lengths of the ellipsoid $\EC\big(\dfrac{\omega}{\omega+\alpha}\big)$, defined in \eqref{Ellips_Strip_Inter}. 
Contrary to all such algorithms in the literature, \cite{Mak:96b,Kur:97,Dur:01,Che:05},
that minimize the size of the ellipsoid $\EC(\beta)$, the optimal value
of $\beta$ chosen here is the one that  fulfills some stability criterion of the
estimation algorithm to be derived, in the manner of \cite{Tan:97,Bec:08,She:18}, 
by minimizing some quadratic measure of the estimation error vector 
in the worst noise case.
 \begin{theo}\label{theo_omega} Let $\EC(\beta)$ given by \eqref{Ellips_Strip_Inter}, where  $\cb\in\Rn$, $\varsigma\in\R_+^*$, $y\in\R$, $P\in~\Rnn$ SPSD  and  $\fb\in\Rn\text{--}\{\zero_n\}$ meet case~4 of Thm \ref{theo_ellips_strip_inter}, then $\breve{\varsigma}(\beta)$ defined in \eqref{sig_ellips_strip_inter} satisfies
\begin{subequations}
\mathl
\begin{align}
& &\breve{\varsigma}(\beta)&=\max_{\bm{x}\in\DC(\breve{\fb},\breve{y})\cap\EC(\cb,\varsigma P)}\Vscr_{\beta}(\x)\label{sigma_maxV}\\
&\text{where }&\Vscr_{\beta}(\x)&\eqg\big(\x-\breve{\cb}(\beta)\big)^T\breve{P}(\beta)^\dag\big(\x-\breve{\cb}(\beta)\big);\label{V_def}
\end{align}
and its minimum is given by
\mathc
\begin{align}\label{omega}
\beta^*& \eqg\arg \min_{\beta\in]0,1[} \breve{\varsigma}(\beta)=
  \begin{cases}
  1-\gamma\abs{\delta}^{-1}  &\text{if } \abs{\delta }>\gamma\\
      0, &\text{otherwise};
      \end{cases}
\end{align}
\end{subequations}
where $\breve{\fb}$, $\breve{y}$, $\alpha$ and $\delta$ are defined in \eqref{breve_f_y_def}, \eqref{alpha_def} and \eqref{delta_def}. 
\end{theo}
{\debdemo 
Applying the generalization of the Sherman-Morrison formula  to the pseudo-inverse of the matrix  \eqref{P_ellips_strip_inter} (\cf Corollary 3.5 \cite{Xu:17}), we can write 
\mathc
\begin{align}
\breve{P}(\beta)^\dag=P^\dag+\frac{\alpha\beta}{1-\beta}{P^\dag P\fb\fb^TPP^\dag}.\label{Pbrevedag}
\end{align}
Since
$\cb\in\range (P)$  (being the center of the ellipsoid of shape matrix $\varsigma P$) and ${\alpha\beta}P\fb\in\range (P)$, by the use of \eqref{c_ellips_strip_inter}, it is clear that $\breve{\cb}(\beta)\in\range (P)$. Now, noticing that $(PP^\dag)^T=P^\dag P$, 
and recalling that, for all $\x\in\range (P)$, $PP^\dag\x=\x$, 
then replacing \eqref{Pbrevedag} in \eqref{V_def} leads to
\mathl
\begin{align}
\Vscr_{\beta}(\x)&\eqg\big(\x-\breve{\cb}(\beta)\big)^T\Big(P^\dag+\frac{\alpha\beta}{1-\beta}{P^\dag P\fb\fb^TPP^\dag}\Big)\big(\x-\breve{\cb}(\beta)\big)\nn\\
&=\big(\x-\breve{\cb}(\beta)\big)^T\Big(P^\dag+\frac{\alpha\beta}{1-\beta}{\fb\fb^T}\Big)\big(\x-\breve{\cb}(\beta)\big).\label{V_2} 
\end{align}
Inserting  \eqref{c_ellips_strip_inter} in \eqref{V_2}, we can show, by the mean of some standard algebraic manipulations, that\footnote{$\Vscr_{\beta}$ is optimized on $\DC(\breve{\fb},\breve{y})\cap\EC(\cb,\varsigma P)$, it is then obvious that $\x\in\range (P)$, since $\x\in\EC(\cb,\varsigma P)$.}, $\forall\x\in\range(P)$,
\begin{align}
\Vscr_{\beta}(\x)
&= \frac{\alpha\beta\gamma^2}{1-\beta}(\breve{y}-\breve{\fb}^T\x)^2-{\alpha\beta\delta^2}
+(\x-\cb)^TP^\dag(\x-\cb),\\
\d\max_{\bm{x}\in\DC(\breve{\fb},\breve{y})}\Vscr_{\beta}(\x)&=\frac{\alpha\beta\gamma^2}{1-\beta}-\alpha\beta{\delta^2}
+(\x-\cb)^TP^\dag(\x-\cb),\\
\ \max_{\bm{x}\in\DC(\breve{\fb},\breve{y})\cap\EC(\cb,\varsigma P)}\Vscr_{\beta}(\x)&=\frac{\alpha\beta\gamma^2}{1-\beta}-\alpha\beta{\delta^2}+\varsigma=\breve{\varsigma}(\beta).
\end{align}
The optimal value of $\beta$ is obtained by zeroing the derivative of $\breve{\varsigma}$: 
\mathc
\begin{align}
\dfrac{\dif\breve{\varsigma}}{\dif\beta}(\beta^{*})=0\eq\gamma^2\left(1-\beta^*\right)^{-2}-\delta^2=0
\eq\beta^{*}=1-\gamma\abs{\delta}^{-1}. 
\end{align}
Since $\beta^*\geq 0$, this solution is conditioned by $\abs{\delta}>~\gamma$; 
if $\abs{\delta}\leq~\gamma$,  the solution to the above minimization problem would be $\beta^*=0$. 
\findemo }
\begin{rema} 
The representation of the output noise vector's bounding set as an intersection of strips, rather than as an ellipsoid, 
enables this optimization problem to have an analytical solution.
\end{rema}
\begin{rema} 
The center, $\breve{\cb }(\beta^*)$, of the (red/small) ellipsoid $\EC(\beta^*)$ is the orthogonal projection of $\cb$, the center of the (blue/big) one, $\EC(\cb ,\varsigma P)$, on the nearest strip boundary (\cf fig. \ref{ell_strip_inters_fig}).
\end{rema}
%
\subsection{ Ellipsoid resulting from the intersection of an ellipsoid with a hyperplane}\label{subsec_hyp}  
\begin{figure}[htbp]
\begin{center}
\includegraphics[angle=0,origin=c,width=0.6\textwidth]{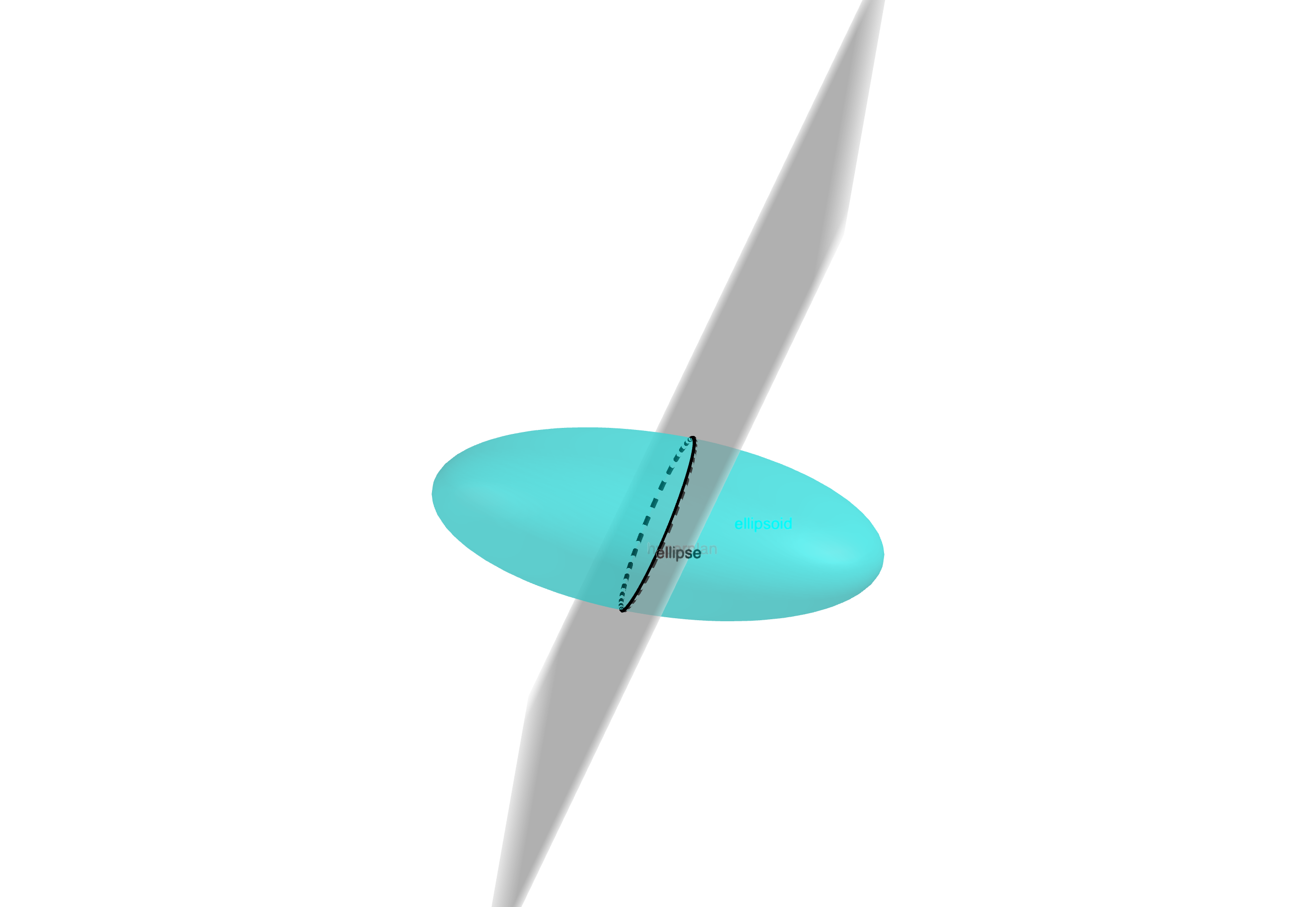}
\caption{Intersection of an ellipsoid with an hyperplane ($n=3$)}
\label{ell_hyp_inters_fig}
\end{center}
\end{figure}

Now let us examine the intersection of an ellipsoid with a hyperplane. This intersection is the projection of the ellipsoid on the subspace represented by this hyperplane and  leads to a degenerate ellipsoid of lesser dimension, whose shape matrix loses one rank with each intersecting (not parallel) hyperplane (\cf fig. \ref{ell_hyp_inters_fig}). 
The theorem below gives the expression of thusly obtained ellipsoid. 
%
\begin{theo}[ellips./hyperplane inters.]\label{theo_ell_hyp_inters}
Let  $\cb\in~\Rn$, $P\in\Rnn$ SPSD, $\varsigma\in\R_+^*$, $\fb\in\Rn\text{--}\{\zero_n\}$ and $y\in\R$,
\begin{subequations}\label{ell_hyp_inters_eq}
\mathc
\begin{align}
\intertext{{if $y>\brho$ or $y<-\urho$,}\hfill\normalfont{(case 1)}}
%
\EC(\cb ,\varsigma P)\cap\HC(\fb,y)&=\emptyset;\label{ell_hyp_empty}\\
\intertext{{else if $y=\brho=-\urho$},\hfill\normalfont{(case 2)}}
\EC(\cb ,\varsigma P)\cap\HC(\fb,y)&=\EC(\cb ,\varsigma P);\label{ell_hyp_ell}\\
\intertext{{otherwise  $(\text{if } -\urho\leq y\leq\brho)$}\hfill\normalfont{(case 3)}}
\EC(\cb ,\varsigma P)\cap\HC(\fb,y)&=\EC(\check{\cb },\check{\varsigma}\check{P}),\label{ell_hyp}\\
\intertext{where }\check{\cb }&\eqg\cb +\alpha\delta{P\fb},\label{eq_x}\\
\check{P}&\eqg P-{\alpha}{P\fb\fb^TP},\label{eq_P}\\ 
\check{\varsigma}&\eqg\varsigma-{\alpha}{\delta^2},\label{eq_sig}\\
\delta&\eqg y-\fb^T\cb \label{eq_delta}
\end{align}
and where $\alpha$, $\urho$  and $\brho$ are defined in \eqref{alpha_def}, \eqref{urho_def} and \eqref{brho_def} \resp.
\end{subequations}
\end{theo}
{\debdemo 
To start with, recall that an affine map 
  $\Fd:~\Rn\rightarrow\Rn$, $\x\mapsto L\x+\bm{a}$ turns an ellipsoid $\EC(\cb,P)$ into another one $\EC(L\cb+\bm{a},L^TPL)$ and the hyperplane $\HC(\fb,y)$ into $\HC(L^{\dag}\fb,y+\fb^TL^{\dag}\bm{a})$. %
Throughout this proof, we'll be changing coordinate systems but dealing with one and the same hyperplane $\HC\eqg\HC(\fb,y)$ and one and the same ellipsoid $\EC\eqg\EC(\cb,\varsigma P)$. 
Consider the vector $\fb\in\Rn-\{\zero_n\}$.\linebreak Two cases (different from those of the Thm) will be distinguished depending on whether $\fb\in{\nul }(P)$ (1) or $\fb\notin{\nul }(P)$ (2).

1. $\fb\in{\nul }(P)$. This means that the matrix $P$ is not SPD but only SPSD (having at least one zero eigenvalue) and  $\fb^TP\fb=0$.  In this case $\EC\subset\HC'$, where $\HC'\eqg\{\x\in\Rn|\fb^T\x=\cb\}$ is the hyperplane of normal vector $\fb$ and containing the center $\cb$ of $\EC$. If $\cb\notin\HC$, \ie, $\fb^T\cb\neq y$ (corresponding to case 1 of the Thm, with $\fb^TP\fb=0$),  $\EC$ is a subset of the hyperplane $\HC'$ parallel to $\HC$ and $\EC\cap\HC=\emptyset$, as in \eqref{ell_hyp_empty}. Otherwise (case 2), $\HC'=\HC$, meaning that $\EC\subset\HC$ and $\EC\cap\HC=\EC$, as in \eqref{ell_hyp_ell}. 

2. Consider now $\fb\notin{\nul }(P)$ and let  $\bar{n}\eqg \rank(P)\le n$. We shall  define the affine transformation  
that maps the unit hypersphere or ball  into the ellipsoid $\EC(\cb,\varsigma P)$: 
\begin{align}
\BC_2^n\xrightarrow{\Fd_1}\EC(\cb,\varsigma P),\text{ \ie, }\Fd_1:\bxb\mapsto\x=\racine{(\varsigma P)}(\bxb+\cb).
\end{align}
Now consider its (pseudo-)inverse transform $\Fd_1^\dag$ that maps the ellipsoid into (a possibly degenerate) unit ball: $\d\EC(\cb,\varsigma P)\xrightarrow{\Fd_1^\dag}~\EC(\bcb,\bP)$ , where
 \begin{align}
 \bcb=\zero_n \text{ and } \bP=\bar{I}_{n,\bar{n}} \text{ where } \bar{I}_{n,\bar{n}}\eqg
 \left[{I_{\bar{n}} \atop 0_{\underb{n}\times \underb{n}} } {0_{ \underb{n}\times \underb{n}} \atop 0_{ \underb{n}\times \underb{n}}}\right],\  \underb{n}\eqg n-\bar{n}
 \label{bc_bP}
\end{align}
and $\d\HC(\fb,y)\xrightarrow{\Fd_1^\dag}~\HC(\bfb,\by)$, \ie, $\bxb\in\HC\eq\bxb^T\bfb=\by$. In the new coordinates system transformed thusly, the unit normal vector to the hyperplane $\HC$ and its minimum signed distance from origin are \resp.  
\begin{align}
\bfb&\eqg\frac{\racine{P}\fb}{\sqrt{\fb^TP\fb}}, \text{ with } \norm{\bfb}=1
\text{ and }\by\eqg\frac{\big(y-\fb^T\cb\big)}{\sqrt{\varsigma \fb^TP\fb}}.\label{bf_by}
\end{align}
 
 \nid Let $\eb\eqg[1\, 0\ldots0]^T$  the first vector of the identity matrix and 
\begin{align}
H\eqg I_n-\tfrac{2}{\norm{\bfb-\eb}^2}(\bfb-\eb)(\bfb-\eb)^T\label{Householder}
\end{align}
is the Householder symmetric ($H=H^T$) and unitary ($HH^T=I_n$) matrix that transforms $\bfb$ into $\eb$: 
$H\bfb=\eb\eq\bfb=H^T\eb=\hb_1$.
 
\nid Next, let 
$\Fd_2:~\bxb~\mapsto~\bbxb~=~H(\bxb~-\by\bfb)$ that transforms the former (second) coordinate system into the third one, in which the considered hyperplane is orthogonal to $\eb$ and contains the origin: $\bbxb\in\HC\eq\bbxb^T\eb=0$, \ie, 
\begin{align}
\HC(\bfb,\by)\xrightarrow{\Fd_2}\HC(\bbfb,\bby)\text{ where }\bbfb\eqg\eb\text{ and }\bby\eqg 0.\label{bbf_bby}
\end{align}
The (possibly degenerate) unit ball $\EC(\bcb,\bP)$ is transformed, by $\Fd_2$, into the (possibly degenerate) hypersphere $\EC(\bbcb,\bbP)$, where 
\begin{align}
\bbcb\eqg H\bcb-\by H\bfb=-\by \eb\text{ and }\bbP\eqg H\bP H^T=\bar{I}_{n,\bar{n}}\label{bbc_bbP}
\end{align}

\nid Now, the distance between the center of the ellipsoid $\EC(\bbcb,\bbP)$ and the hyperplane $\HC(\bbfb,\bby)$, $\abs{\bbcb^T\bbfb-\bby}=\abs{-\by\eb^T\eb-0}$ $=\abs{\by}$ is compared to the projection of the radius of the former onto the normal vector to the latter:
\begin{align}
\sqrt{\bbfb^T\bbP\bbfb}=\sqrt{\eb^T\bar{I}_{n,\bar{n}}\eb}=1.
\end{align}
If $\abs{\by}>1$ (case 1 with $\fb^TP\fb\neq0$), then $\EC\cap\HC=\emptyset$. Otherwise (case 3),
the spheroid\footnote{A  spheroid is a possibly degenerate hypersphere.}  
resulting from the intersection of the (possibly degenerate) hypersphere $\EC(-\by\eb,\bar{I}_{n,\bar{n}})$ and the hyperplane $\HC(\eb,0)$ is $\EC(\bbccb,\bbcP)$ where 
\begin{subequations}\label{bbcc_bbcP}
\begin{align}
\bbccb&\eqg\bbcb+\eb^T\bbcb\eb= -\by\eb+(\eb^T\eb)\by\eb=\zero_n,\label{bbcc}\\
\bbcP&\eqg\big(1-(\eb^T\bbcb)^2\big)\big(\bar{I}_{n,\bar{n}}-\eb\eb^T\big)=\big(1-\by^2\big)\big(\bar{I}_{n,\bar{n}}-\eb\eb^T\big)\label{bbcP}.
\end{align}
\end{subequations}

This ellipsoid is expressed in the third coordinate system. Well, we have to find its expression in the orignal one and for this purpose, the inverse former transformations will be applied in reverse order: $\EC(\bbccb,\bbcP)\xrightarrow{\Fd_1\circ\Fd_2^{-1}}~\EC(\ccb,\csig\cP)$.
To start with, we'll apply the inverse transformation $\Fd_2$ to the spheroid:\linebreak $\EC(\bbccb,\bbcP)\xrightarrow{\Fd_2^{-1}}\EC(\bccb,\bcP)$, to obtain 
\begin{subequations}\label{bcc_bcP}
\begin{align}
\bccb&\eqg H^T\bbccb+\by\bfb=\by\bfb\\
\bcP&\eqg H^T\bbcP H=\big(1-\by^2\big)\big(H^T\bar{I}_{n,\bar{n}}H-H^T\eb\eb^TH\big)=\big(1-\by^2\big)\big(\bar{I}_{n,\bar{n}}-\bfb\bfb^T\big).\nn
\end{align}
\end{subequations}

\nid Then, applying $\Fd_1$: 
$\EC(\bccb,\bcP)\xrightarrow{\Fd_1}~\EC(\ccb,\csig\cP)$, yields to 
\begin{subequations}\label{cc_cP}
\begin{align}
\ccb&\eqg \racine{(\varsigma P)}\bccb+\cb=\cb+\racine{(\varsigma P)}\by\bfb\\
\csig\cP&=\racineT{(\varsigma P)}\bcP\racine{(\varsigma P)}=\varsigma\big(1-\by^2\big)\big(P-\racine{P}\bfb\bfb^T\racine{P}\big).
\end{align}
\end{subequations}
Lastly, choosing $\csig\eqg\varsigma\big(1-\by^2\big)$ and $\cP\eqg P-\racine{P}\bfb\bfb^T\racine{P}$ and replacing afterwards $\by$, $\bfb$, $\ccb$, $\cP$ and $\csig$ by their respective expressions, \eqref{bf_by} and \eqref{cc_cP},  we get to \eqref{eq_P}$-$\eqref{eq_sig}.
 \findemo
  }
\subsection{ The overall state estimation algorithm}\label{sec_algo}
\begin{theo}\label{theo_meas_update}
Let us set the following assignments 
\begin{subequations}\label{ellips_correc}
\mathc
\begin{align}
\byki&\leftarrow \min(\byki,\brhoki)\text{ and }
\uyki\leftarrow \max(\uyki,-\urhoki).\label{by_uy_redef}\\
\intertext{If $\xka$  satisfies \eqref{system0}  
meeting \eqref{bounds}, then }
\xka\in\SC_k\subseteq\EC_k&\eqg\EC(\xk ,\sigk P_k),\\
\intertext{where $\SC_k$ is defined in \eqref{SC_set} and $\forall k\in\N^*$, }
\sigk&\eqg\varsigma_{k_{p_k}}, \Pk\eqg P_{k_{p_k}} \text{ and }\xk\eqg\xe_{k_{p_k}};\label{eq_correc}\\
 \sigkz&\eqg\sigkkk, \Pkz\eqg\Pkk \text{ and }\xkz\eqg\xkk;\label{eq_correc0}\\
\intertext{$\xkk\text{ and }\Pkk$  are given in \eqref{x_pred} and \eqref{predic_eq_glo}; and for $i\in\{1,\ldots,p_k\}$, }  
\Pki&\eqg\Pkiii-{\alphak{i}\betak{i}}{\phibki \phibki ^T},\label{eq_correc-P}\\ 
\xki&\eqg\xkiii+{\alphak{i}\betak{i}}\delki \phibki, \label{eq_correc-x}\\ 
\sigki&\eqg\sigkiii-{\alphak{i}\betak{i}^2\delkis};\label{eq_correc-sig}\\
\intertext{where }
\alphak{i}&\eqg 
\begin{cases}
\thetaki^{-1} , &\text{if }p_k\neq 0 \text{ and }  \lamki  \neq 0,\\ 
0, &\text{otherwise};\end{cases}\label{eq_correc-alph}\\
\betak{i}&\eqg \begin{cases}
1,&\text{if }\uyki=\byki\text{ and }-\urhoki\neq\brhoki,\\ 
1-\gammaki\abs{\delki}^{-1}, &\text{else if }\abs{\delki }>\gammaki, \\
&\text{and }(-\urhoki<\uyki\text{ or }\byki<\brhoki),\\ 
 0,&\text{otherwise;}
  \end{cases}\label{eq_correc-bet}\\
   \delki&\eqg   \tfrac{1}{2}({\byki+\uyki}-\brhoki+\urhoki),\label{eq_correc-del}\\
   \gammaki&\eqg  \frac{1}{2}(\byki-\uyki),\label{eq_correc-gam}\\
   \thetaki&\eqg \fbki^T\phibki, \label{eq_correc-the}\\
\phibki &\eqg P_{k_{i-1}}\fbki,\label{eq_correc-phi}\\
\brhoki&\eqg \lamki+\fbki^T\xkiii 
\text{ and }
\urhoki \eqg
2\lamki-\brhoki,\label{eq_correc-brho-urho}\\
 \lamki&\eqg(\sigkiii\thetaki)^{\frac{1}{2}}.\label{eq_correc-lam}
\end{align}
\end{subequations}
\end{theo}
{\debdemo 
Direct application of Thms \ref{theo_ellips_HSpace_inter}, \ref{theo_ellips_strip_inter}, \ref{theo_omega} and \ref{theo_ell_hyp_inters} to $\SC_k$ given in \eqref{SC_set}.
\findemo }

The time prediction stage given by Thms \ref{theo_time_update} and \eqref{predic_eq_glo} and the measurement correction phase, given by Thm \ref{theo_meas_update} are concatenated to form the hole state estimation algorithm presented in Algorithm \ref{algo}, where  $N$ is the number of samples. 
{%
\begin{algorithm}
\caption{Computation of the ellipsoid $\EC(\xk,\sigk P_{k})$}
\begin{algorithmic}[1]\label{algo}
	\REQUIRE $\xz$, $\sigz$, $P_0$, $N$ 
\ENSURE $\xk$, $\sigk$, $P_{k}$
\STATE $n\leftarrow$ size of $\xz$\;
\FOR{$k=1,2,\ldots, N$}
\STATE {\bf Input:}
$A\kkk$,  $B\kkk$, $R\kkk$, $\taubkk$, intervening in \eqref{system0},
\newline 
${F}_k=[\fbki]_{i\in\{1,\dots,p_k\}}$, $\bybk=[\byki]_{i\in\{1,\dots,p_k\}}$, $\uybk=[\uyki]_{i\in\{1,\dots,p_k\}}$, as in \eqref{bounds};
\LCOMMENT{\bf // Time prediction //}
\STATE $\Pkkz\leftarrow\Akk^{T}\Pkkk \Akk$; \COMMENT{\bf Initialization} 
\STATE  $\bar{\mu}_{{k-1}_0}\eqg \sqrt{\sigkkk\tr(\Pkkz)}$; 
$\d\bar{\mu}_{k-1}\eqg\sum_{i=1}^m{\norm{\rkki}}$ as in \eqref{mu0_def} and \eqref{mu_def};
\STATE  $\d\bar{R}_{k-1}\eqg\sum_{i=1}^m\norm{\rkki}^{-1}\rkki\rkki^T$, as in \eqref{bR_def};
\STATE    $\Pkk\eqg\big(1+\tfrac{\bar{\mu}_{k-1}}{\bar{\mu}_{k-1_0}}\big)\big(\Pkkz+\tfrac{\bar{\mu}_{k-1_0}}{\sigkkk}\bar{R}_{k-1}\big)$ as stated in \eqref{predic_eq_glo};  
\STATE  $\xkk\eqg\Akk\xkkk+B_{k-1}\taubkk$ conforming to \eqref{x_pred}; 
\LCOMMENT{\bf // Measurement correction //}
\STATE $p_k\leftarrow $ number of columns of $F_k$;
\IF{$p_k=0$}
\STATE $\xk\leftarrow\xkk$; $\Pk\leftarrow \Pkk$; $\sigk\leftarrow\sigkkk$; 
\ELSE
\STATE $\sigkz\leftarrow~\sigkkk$; $\Pkz\leftarrow\Pkk$; $\xkz\leftarrow~\xkk$; \COMMENT{\bf Initialization} 
\FOR{$i=1,\ldots p_k$}
\STATE  $\phibki \eqg P_{k_{i-1}}\fbki$; $\thetaki\eqg \fbki^T\phibki$ as in  \eqref{eq_correc-phi} and \eqref{eq_correc-the};
\IF{$\thetaki=0$}
\STATE $\xk\leftarrow\xkk$; $\Pk\leftarrow \Pkk$; $\sigk\leftarrow\sigkkk$; 
\ELSE 
\STATE  $\alphak{i}\eqg \thetaki^{-1}$;  $\lamki\eqg(\sigkiii\thetaki)^{\frac{1}{2}}$ as in \eqref{eq_correc-alph}, \eqref{eq_correc-lam};
\STATE $\brhoki\eqg \lamki+\fbki^T\xkiii $; $\urhoki \eqg2\lamki-\brhoki$,
as in  \eqref{eq_correc-brho-urho}; 
\STATE $\byki\leftarrow \min(\byki,\brhoki)$;
$\uyki\leftarrow \max(\uyki,-\urhoki)$ as in  \eqref{by_uy_redef};
\STATE  $\delki\eqg   \tfrac{1}{2}({\byki+\uyki}-\brhoki+\urhoki)$; as in \eqref{eq_correc-del};
\STATE $\gammaki\eqg  \frac{1}{2}(\byki-\uyki)$, as in \eqref{eq_correc-gam}; 
\IF{$\uyki=\byki\text{ and }-\urhoki\neq\brhoki$}
\STATE $\betaki = 1$;
\ELSIF{$\abs{\delki }>\gammaki$}
\STATE $\betaki = 1-\gammaki\abs{\delki}^{-1}$;
\ELSE 
\STATE $\betaki = 0$;
\ENDIF
\STATE  $\Pki\eqg\Pkiii-{\alphak{i}\betak{i}}{\phibki \phibki ^T}$, as in \eqref{eq_correc-P};
\STATE $\xki\eqg \xkiii+{\alphak{i}\betak{i}}\delki \phibki $, as in \eqref{eq_correc-x};
\STATE $\sigki\eqg \sigkiii-{\alphak{i}\betak{i}^2\delkis}$ as in \eqref{eq_correc-sig};   
\ENDIF
\ENDFOR
\STATE $\xk\leftarrow\x_{k_{p_k}}$; $\Pk\leftarrow P_{k_{p_k}}$; $\sigk\leftarrow\varsigma_{k_{p_k}}$; 
\ENDIF
\ENDFOR
\end{algorithmic}
\end{algorithm}
}
 \begin{rema}
 In the case where $\hscrk\neq\emptyset$, the matrix $\Pki$ 
 loses rank  with each intersecting hyperplane $\HC\ki$, $i\in\hscrk$, thusly entailing the progressive flattening of the ellipsoid $\EC_{k_i}$. Depending on the rank of the matrix $R_k$ (on which no assumption is made), the rank of $P_{k+1/k}$ can be recovered at the time-update phase. 
\end{rema}
\begin{rema}
Setting either $\alphak{i}=0$ or $\betak{i}=0$ results in freezing $\EC_{k_{i-1}}$, meaning that the corresponding measurements $\fbki,\uyki,\byki$ do not bring any useful information.
\end{rema}
 \begin{rema}
 The cases 1 of Thms  \ref{theo_ellips_HSpace_inter}, \ref{theo_ellips_strip_inter} and \ref{theo_ell_hyp_inters} are not explicitly treated in this theorem assuming that they can not occur since the intervening measurements should be consistent with the system model;
yet the case where the measurement $\fbki,\uyki,\byki$ is aberrant is implicitly considered,
setting again either $\alphak{i}=0$ or $\betak{i}=0$,  preventing  so the updating of the ellipsoid $\EC_{k_{i-1}}$.
 \end{rema}
\begin{rema}
This algorithm is of low computational complexity. Indeed, all the operations are simple sums and products: they were optimized in this regard and are thence suitable for systems with high dimensional state vector (bif $n$) and with many measurements (big $p_k$). The intermediate variables $\alphaki$, $\thetaki$, $\lamki$, $\phibki$ were added on to perform redundant vector and matrix operations only once. Thereby noticing that $\fbki^T\xkiii=\tfrac{1}{2}(\brhoki-\urhoki)$ allows to determine $\delki\eqg\tfrac{1}{2}({\byki+\uyki})-\fbki^T\xkiii$ and $\urhoki\eqg\lamki-\fbki^T\xkiii$ using addition of scalars, in \eqref{eq_correc-del} and \eqref{eq_correc-brho-urho} \resp., rather than multiplication of possibly high dimensional vectors.
\end{rema}
\begin{rema}
For more numerical stability and in order to avoid the explosion of the matrix $P_k$, caused by the set summations at the prediction step, the assignments \eqref{eq_correc0} can be replaced by $\Pk\leftarrow \frac{\varsigma_{k_{p_k}} }{\sigz}P_{k_{p_k}}$, 
$\bar{\varsigma}_{k}\leftarrow\frac{\sigz\varsigma_{k_{p_k}} }{\bar{\varsigma}_{k-1}}$ and $\sigk\leftarrow\varsigma_{0}$. 
Then $P_k$ would, by itself, represent the shape of the ellipsoid $\EC_k$ up to a constant factor $\sigz^{-1}$ and the new variable $\bar{\varsigma}_{k}$ is introduced to keep track of the decreasing parameter $\sigk$, with $\bar{\varsigma}_{0}=\sigz$.
\end{rema}
\section{{Algorithm properties and stability analysis}}
The proposed algorithm is designed in such a way as to fulfill the requirements \ref{requirement1} - \ref{requirement3}, expressed in the $\mathsection$\ref{sec_prob_form} and this is what will be shown in this section.
%
The stability requirement \ref{requirement3} exploits the Input-to-State stability concept: roughly speaking, for an ISS system, 
inputs that are bounded, \guil{eventually small}, or convergent, should lead to the state vector with the respective property; and that the $\zero$-input system 
should be globally asymptotically stable. More formal definitions and results are given in the Appendix \ref{sec_ISS}.

\begin{minipage}{\textwidth}
\begin{theo}\label{theo_prop}
Consider the system \eqref{system0} subject to \eqref{bounds} and its state estimation algorithm given by Thms \ref{theo_time_update} and \ref{theo_meas_update}.
\begin{enumerate}
  \item\label{theo_prop_E0_Ek} If $\xza\in\EC(\xz,\sigma_0^2 P_0)$, then $\forall k\in\N^*$, $\xka\in\EC(\xk,\sigk
  P_k)$;
  \item\label{theo_prop_acceptable} The vector $\xk$ is \emph{acceptable}, \ie, it  satisfies \eqref{boundb_HSpace}--\eqref{bound_Strip}: 
  \begin{align}
  \forall k\in\KC,\ \xk\in\bigcap_{i\in\gscrk}\GC\ki\cap\bigcap_{i\in\dscrk}\DC\ki\cap\bigcap_{i\in\hscrk}\HC\ki,  
  \end{align}
  where $\KC\eqg\{k\in\N|p_k\neq 0\}$.
  \item\label{theo_prop_lim_sig} The sequence
  $\left(\sigk\right)_{k\in\N^*}$ 
   is decreasing and convergent on $\R_+$, where $\sigk$ is defined in \eqref{eq_correc-sig} and it satisfies
   \begin{subequations}\label{Lyap_sig}
   \mathc
  \begin{align}
  \sigk&=\max_{\xb\in\SC_k} \Vscr_{k}(\x),  \text{ where  $\SC_k$ is given by \eqref{SC_set} and}\nn\\
 \Vscr_{k}(\xb)&\eqg\big(\x-\xk\big)^T{P}_k^\dag\big(\x-\xk\big). \label{Lyap_def}
  \end{align}
   \end{subequations}
\end{enumerate}
\end{theo}
\end{minipage}
{\debdemo 
\begin{enumerate}
\item 
This point is satisfied by construction. Indeed, from \eqref{SC_set}, Thms \ref{theo_time_update} and \ref{theo_meas_update}, we have
\setcounter{Thmnbr}{\value{enumi}}
\begin{align}
\xza\in\EC_0&\Rightarrow\left(\x_1\in\EC_{1/0}\right)\land\bigg(\x_1\in\big(\PC_1\cap\ZC_1\cap\bigcap_{i\in\hscr{1}}\HC_{1_i}\big)\bigg)\nn\\
&\Rightarrow\x_1\in\SC_{1}\Rightarrow\x_1\in\EC_{1}\Rightarrow\cdots
\Rightarrow\xkka\in\EC\kkk\nn\\
&\Rightarrow\left(\xka\in\EC\kk\right)\land\bigg(\xka\in\PC_k\cap\ZC_k\cap\bigcap_{i\in\hscrk}\HC_{k_j}\bigg)\nn\\
&\Rightarrow\xka\in\SC_{k}\Rightarrow\xka\in\EC_{k}, \ \forall k\in\N^*.
\end{align}
%
%
\item 
This point is also granted by construction. To check it, consider \linebreak $\xkz\eqg\xkk$.  From \eqref{eq_correc-x},
\mathc
\begin{align*}
\fbk{i}^T\xki=\fbk{i}^T\xkiii+{\alphak{i}\betak{i}}\delki\fbk{i}^T\phibki .
\end{align*}
If $\fbk{i}^T\phibki =0$ or $\abs{\delki}\leq \gamki$, it means that $\xki$ is already in $\DC\ki$ or $\HC\ki$. 
Else, 
\begin{align}
\fbk{i}^T\xki=\fbk{i}^T\xkiii+\betak{i}\delki.\label{pr_acc1}
\end{align}
Now, if $i\in\hscrk$, $\beta=1$ according to \eqref{eq_correc-bet}, then inserting \eqref{eq_correc-del} in \eqref{pr_acc1}, results in $\fbk{i}^T\xki=\yki$ meaning that $\xki\in\HC\ki$. Otherwise, $\beta=1-~\!\gammaki\abs{\delki}^{-1}$ and  $\fbk{i}^T\xki-\yki=-1$, if $\delki<-\gammaki$ and $\fbk{i}^T\xki-\yki=1$, if $\delki>\gammaki$; this means that $\xki\in\DC\ki$. Combining these results for $i\in\gscrk\cup\dscrk$, leads to $\xk\in\SC_k$, where $\SC_k$ is defined  in \eqref{SC_set} and considering \eqref{bounds}, the proof of  the point \ref{theo_prop_acceptable} is achieved.
\item 
From \eqref{eq_correc-sig} of Thm \ref{theo_omega}, $\sigki-\sigkiii=-{\alphak{i}\betak{i}^2\delkis}$. Since $\alphak{i}$, defined in \eqref{eq_correc-alph}, is a quadratic form when it is non-zero, it is obvious that $\sigki-\sigkiii\leq 0$. From \eqref{eq_correc} and \eqref{eq_correc0},
$\sigk\eqg\varsigma_{k_{p_k}}$ and  $\sigkkk\eqd\sigkz$, then 
\begin{align}
\sigk-\sigkkk=\varsigma_{k_{p_k}}-\sigkz=-\sum_{i=0}^{p_k}{\alphak{i}\betak{i}^2\delkis}\leq0.
\end{align}
The sequence $\big(\sigk\big)_{k\in\N}$ is decreasing, bounded above by $\sigz$ and hence convergent.\findemo
\end{enumerate}
 }
%
\begin{theo}\label{theo_ISS} Let
\begin{subequations}\label{bF_bp_KC_def}
\begin{align}
\bar{F}_{k}&\eqg{F}_{k}(\gscrk\cup\dscrk)\eqg[\fbki]_{i\in\gscrk\cup\dscrk}\in\R^{n\times\bpk },\label{bF_def}\\
\bpk &\eqg\card(\gscrk\cup\dscrk)\text{ and }\bar{\KC}=\{i\in\N^*|\bpk \neq 0\}.\label{bp_KC_def}
\end{align}
\end{subequations}
If the pairs $\{A_{k},\bar{F}_{k}^T\}$ 
and $\{A_{k},R_k\}$  are \emph{sporadically observable}\footnote{\cf Definition \ref{defi_obs_spora} in the Appendix \ref{sec_obs_contr}}  and completely uniformly  controllable \resp.,
then
\begin{enumerate}
     \item\label{theo_prop_vol} the volume of $\EC_k$ and all its axes lengths are bounded; 
    %
    \item\label{theo_prop_ISS} $\Vscr_{k}$, given in \eqref{Lyap_def}, is an ISS-Lyapunov function for the estimation error $\dxbk\eqg\xka-\xk$, which is ISS\footnote{\cf Definitions \ref{defi_ISS}, \ref{defi_ISS_Lyap} and the Lemma \ref{lemm_ISS}.}. 
%
\end{enumerate}
\end{theo}
{\debdemo 
The proof is detailed in the Appendix \ref{sec_stability}.
\findemo }
\section{\uppercase{Numerical simulations}}

\begin{figure}[htbp]
\centering
     \begin{subfigure}[b]{0.49\textwidth}
         \centering
         \includegraphics[width=\textwidth]{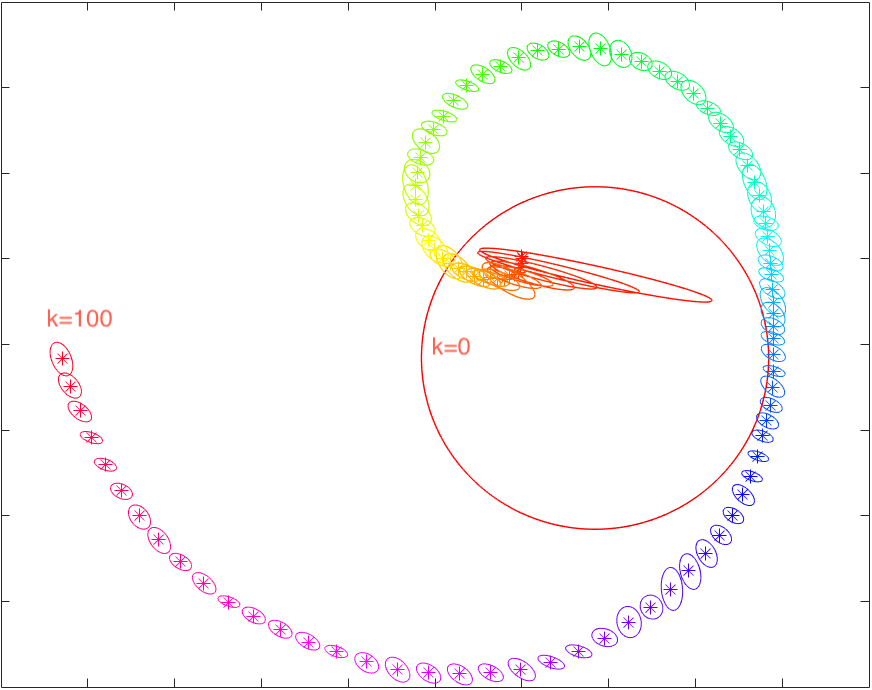}
         \caption{Stable state model}
         \label{stable_fig}
     \end{subfigure}
     \hfill
     \begin{subfigure}[b]{0.49\textwidth}
         \centering
         \includegraphics[width=\textwidth]{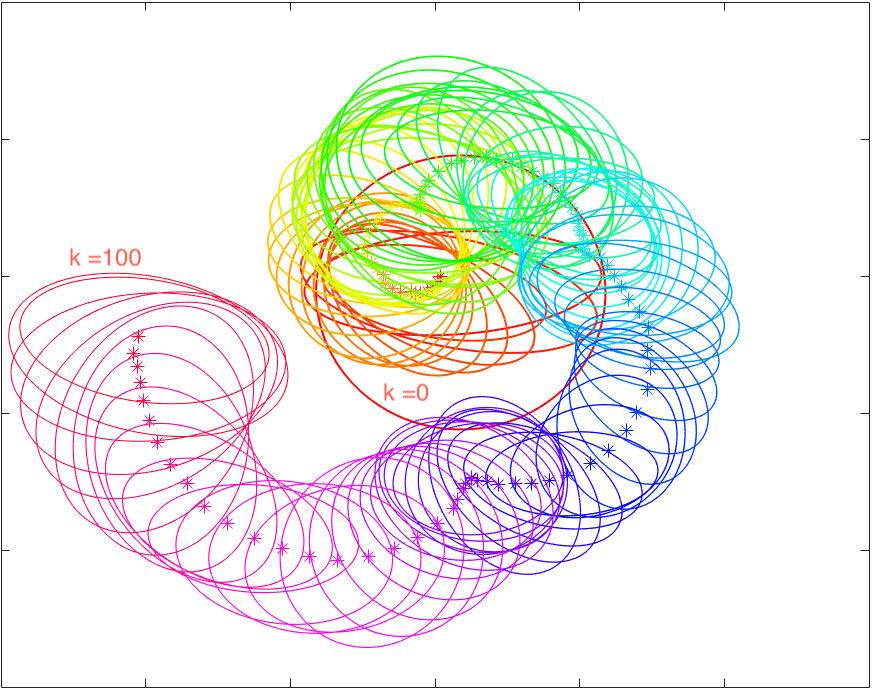}
         \caption{Unstable state model}
         \label{unstable_fig}
     \end{subfigure}
     \hfill
\end{figure}

First, for the sake of graphic illustration, the presented algorithm is applied to a second order randomly generated system with coil-shaped input and one (either strip or halfspace type) measurement, also randomly generated at each time-step for $k=0,\cdots,100$. The figures \ref{stable_fig} and  \ref{unstable_fig} show the evolution of the ellipsoid $\EC_k$ for a stable model (the eigenvalues of state matrix $A\eqg A_k$ are less than 1) and a model at the stability limit (the eigenvalues $A\eqg A_k$ are 1), \resp.

Secondly, in order to evaluate the algorithm performances, the matrices of the system model \eqref{system0} and \eqref{bounds} are generated randomly for two values of the state dimension: $n=10$ and $n=~100$, with $\varrho=q=r=s=\frac{n}{2}$, $m=n$, $\pi=\frac{n}{5}$ and $\mub=\mub^{\tr}_k$; the input vector $B_{k-1}\taubkk$ contains sine entries of random magnitude and frequency. $P_0=100I_n$, $\sigz=1$ and $\xkz$ randomly chosen on the boundary of $\EC_0$. 
The measurements are available at all time steps $\KC=\{1,\ldots,N\}$ in the case 1; at some randomly chosen time steps, in case 2 and $\KC=\emptyset$ in the case 3 (where only prediction stage is performed without any measurement correction).
For each case, the simulations are run 25 times under MATLAB R2018b on Intel Core i7 (2.3GHz, 8G RAM), each one for a different system model and containing $N=~100$ time steps. The results are summarized in Table \ref{Table}.
Let  $\overline{\varsigma\tr(P)}\eqg\underset{k\in\{1,\ldots,N\}}{\mean\sigk\tr(P_k)}$: the average sum of $\EC_k$'s squared axes lengths, 
$\overline{\norm{\tilde{\xb}}}\eqg\underset{k\in\{1,\ldots,N\}}{\mean\norm{\dxbk}}$: the mean estimation error vector norm and $T$: the average computational time for the simulation horizon of $N$ time steps. 
\begin{table}[ht]
\centering
\begin{tabular}[t]{|c|c|c|c|c|c|c|}
\hline
$n$&$\KC$ & $\dfrac{\sigz \tr(P_N)}{\varsigma_N \tr(P_0)}$& $\overline{\varsigma\tr(P)}$ & $\dfrac{\norm{\tilde{\xb}_N}}{\norm{\tilde{\xb}_0}}$& $\overline{\norm{\tilde{\xb}}}$ &$T(ms)$\\ \hline
\multirow{3}{*}{$10$}&case 1&0.010&1\,141&0.027&3.02&18\\ \cline{2-7}
&case 2&0.026&2\,206&0.053&4.60&15\\ \cline{2-7}
&case 3&0.058&7\,540&0.066&7.25&10\\ \hline
\multirow{3}{*}{$100$}&case 1&0.712&7$\cdot10^5$&0.264&26.68&1870\\ \cline{2-7} 
&case 2&0.707&2$\cdot10^6$&0.269&46.19&1321\\ \cline{2-7}
&case 3&5.774&5$\cdot10^6$&0.702&71.16&700\\ \hline
\end{tabular}
\caption{Simulation results}
\label{Table}
\end{table}
It is plain to see that the algorithm considered here exhibits better performances for systems of rather smaller dimension but it can still be fairly efficiently used with very high dimensional systems provided enough measurements are available. Moreover, given its low running time, it can be implemented online with such systems.  
\section{\uppercase{Conclusion}}
We have proposed an ellipsoidal state characterization for discrete-time linear dynamic models with linear-in-state sporadic measurements, which are corrupted by additive unknown process and measurement disturbances, enclosed by zonotopes, on one hand and subject to linear equality and inequality constraints on the other hand.
Here is a turnkey, ready to use, easily implementable algorithm, without any parameter to tune. 

A particular attention was accorded first to the stability of the estimation algorithm, which is ISS despite of the irregularity of the measurements; then to its computational efficiency. Indeed, the proposed algorithm is composed of only low demanding, optimized in this sense operations (matrix sums and products), no costly tools nor heavy operations such as interval arithmetic or LMI, not even matrix inversion have to be performed,
what makes this algorithm suitable for high dimensional systems. 
Furthermore, the challenge faced in other Kalman-like algorithms, inherent to the inversion of the state error covariance matrix
--which is inevitably singular in presence of equality constraints--is circumvented here since the matrix $P_k$  is actually never inverted.
\appendix
\section{Appendix: Stability analysis}\label{sec_stability}
%
\subsection{Kalman filter analogy}\label{sec_KF}
To prove Thm \ref{theo_ISS}, we'll be using the observability and controllability properties of the Kalman filter. For this purpose, we have to show the analogy of the latter with the proposed algorithm. 
Consider  the following  linear time-varying stochastic system with some bounded matrix $\bar{A}_k\in\Rnn$:
\mathc
\begin{subequations}\label{kf_sys}
\begin{align}
\xibk&=\bar{A}\kkk\xibkk+B\kkk\taubkk+R\kkk\wkk,\  k\in\N^*\\
\cybk&=\bar{F}_{k}^T\xibk+\vk,\ \forall k\in\bar{\KC},\text{ (\cf\eqref{bp_KC_def})}\label{kf_y_def}\\
\text{where } \wk&\sim\NC(\zero_n,W_k)\\
  \vk&\sim\NC(\zero_{\bpk },V_k)
\end{align}
\end{subequations}%
where $\xibk\in\Rn$ is the unknown state vector, $\cybk\eqg[\yki]_{i\in\gscrk\cap\dscrk}\in\R^{\bpk }$,\linebreak $\ybk\eqg\tfrac{1}{2}(\bybk-~\!\uybk)$ and $\bar{F}_{k}\in~\R^{n\times\bpk }$, defined in \eqref{bF_bp_KC_def}, are the output vector and the observation matrix \resp.; 
$B\kkk\taubkk$ is the known input intervening in \eqref{state_eq0};
and $\wkk$ and $\vk$ are gaussian centred noise vectors of covariance matrices $W\kkk$ and $V_k$ resp. 
Now consider the Kalman filter, designed for the system~\eqref{kf_sys}:
\begin{subequations}\label{kf}
\begin{align}
 \xibhk   & =  \xibhkk+K_{k}\delk   \label{kf_xest}        \\
   \bar{P}_{k} & = (I_n-K_{k}\bar{F}_{k}^T)\bar{P}\kk    \label{kf_Pest}\\
   \delk & \eqg  \cybk-\bar{F}_{k}^T\xibhkk  \label{kf_innovation}\\
    K_{k} & \eqg   \begin{cases}
   \bar{P}\kk \bar{F}_{k}(\bar{F}_{k}^T\bar{P}_{k-1}\bar{F}_{k} + V_k)^{-1},&\text{if } k\in\bar{\KC}\\
      0_{n\times\bpk }, &\text{otherwise};
      \end{cases}
 \label{kf_gain}\\
    \xibhkk &=\bar{A}\kkk\xibhkkk+B\kkk\taubkk\label{kf_xpred}\\
\bar{P}\kk&=\bar{A}\kkk \bar{P}\kkk\bar{A}\kkk^T+R\kkk W\kkk R\kkk^T.\label{kf_Ppred}
\end{align}
\end{subequations}
The time prediction stage, \eqref{predic_eq}, 
of Thm \ref{theo_time_update} 
can be seen as the prediction stage of the Kalman filter \eqref{kf_xpred}-\eqref{kf_Ppred} and the measurement correction stage  \eqref{ellips_correc}, given in Thm \ref{theo_meas_update} is nothing else than \eqref{kf_xest}-\eqref{kf_gain}. This is stated in Proposition \ref{algo_eq_kf}. 
Forasmuch as the Kalman filter undergoes numerical stability issues 
when the system \eqref{kf_sys} is subject to equality constraints (the matrix $\bar{F}_{k}^TP_{k-1}\bar{F}_{k} + V_k$ in the Kalman gain, \eqref{kf_gain}, becoming ill-conditioned), \eqref{bound_HyperP0} are not considered for the moment.
\begin{propo}\label{algo_eq_kf}
If $\xk$ is computed in line with the algorithm composed of \eqref{x_pred}, \eqref{predic_eq_glo}  
and \eqref{ellips_correc} 
and if $\xibhk$ is the Kalman estimator \eqref{kf} designed for the system \eqref{kf_sys},  such that
\mathc
\begin{subequations}\label{kf_A_W_V_def}
\begin{align}
\bar{A}_k&\eqg\lambda_k A_k,\ \lambda_k\eqg\sqrt{1+\tfrac{\bar{\mu}_{k}}{\bar{\mu}_{k_0}}}, \label{kf_A}\\
W_k&\eqg   \tfrac{\bar{\mu}_{k}+\bar{\mu}_{k_0}}{\sigk}\diag(\bar{\mu}_{k_i}^{-1})_{i\in\{1,\cdots,m\}},  \label{kf_W} \\
 \hspace{-0mm} V_k&\eqg \diag\big(\omki^{-1}\big)_{i\in\gscrk\cap\dscrk},\text{ where }  \omki\eqg\tfrac{\alphak{i}\betak{i}}{1-\betak{i}} \label{kf_V}
 \end{align}
 \end{subequations}
 and where $\bar{\mu}_{k_0}$, $\bar{\mu}_{k}$, $\alphaki$, $\betaki$ are defined in \eqref{mu0_def}, \eqref{mu_def}, \eqref{eq_correc-alph}, \eqref{eq_correc-bet} \resp.
 \mathl
\begin{align}
 \sigk &\eqg \sigkkk-\delkT\Upsilon_k\delk,  \text{ where }  \Upsilon_k \eqg\diag\big(\alphak{i}\betak{i}^2\big)_{i\in\gscrk\cap\dscrk}, \text{ for a fixed $\sigkz$; }\label{kf_sig}  
\end{align}
and if $\xz=\xibh_0$ and $\bar{P}_0=P_0$, then
\mathc
\begin{align}
\forall k\in\N^*,\ \xk= \xibhk \text{ and } \bar{P}_k=P_k.
\end{align}
\end{propo}
\debdemo 
Replacing $\betak{i}={(\omki+\alphak{i})^{-1}}{\omki}$ and $\alphak{i}$ from \eqref{eq_correc-alph} in \eqref{eq_correc-P}, the latter can be rewritten 
\mathc
\begin{align}
\Pki&=\Pkiii-{\Pkiii\fbki \big(\fbki ^T\Pkiii\fbki +\omki^{-1}\big)^{-1}\fbki ^T\Pkiii}.\nn
\end{align}
Then, using the inversion lemma, 
it comes that 
\begin{align}
P_k^{-1}=P_{k_{\bpk }}^{-1}=P_{k_{0}}^{-1}+\sum_{i=1}^{\bpk }\omki{\fbki \fbki ^T}. 
\end{align}
Recalling that $P_k=P_{k_{\bpk }}$ and that $P_{k_0}=\Pkk$ and noticing that  
\begin{align}
\sum_{i=1}^{\bpk }{\omki\fbki \fbki ^T}&=\bar{F}_{k}^TV_k\bar{F}_{k}\\
\intertext{we have}
P_k^{-1}&=\Pkk^{-1}+\bar{F}_{k}^TV_k^{-1}\bar{F}_{k}.\label{pr_acc3}
\end{align}
Applying the inversion lemma again to \eqref{pr_acc3}, the algorithm \eqref{ellips_correc} can be rewritten as \eqref{kf_y_def}, \eqref{kf} and \eqref{kf_A_W_V_def}.
Finally, 
using $P\kkp$ defined in \eqref{predic_eq_glo}
and considering \eqref{kf_A} and \eqref{kf_W},
we obtain \eqref{kf_xpred}-\eqref{kf_Ppred}, which completes the proof.
\findemo 
\subsection{Observability and controllability}\label{sec_obs_contr}
Before examining the observability and the controllability of the studied system \eqref{system0}-\eqref{bounds0}, we need to define  the controllability and observability gramians, of length $l\in\N$: 
 \begin{subequations}\label{gramian_con_obs}
 \mathc
\begin{align}
\bar{\CC}_{k+l,k}&\eqg\sum_{i=k}^{k+l-1}\bar{\lambda}_{i+1,k}^{-2}{\Phi}_{k,i+1}{R}_iW_i{R}_i^T{\Phi}_{k,i+1}^T\label{gramian_con}\\
\bar{\OC}_{k+l,k}&\eqg\sum_{i=k}^{k+l}\bar{\lambda}_{i,k}^{2}{\Phi}_{i,k}^T\bar{F}_{i}V_i^{-1}\bar{F}_{i}^T{\Phi}_{i,k}\label{gramian_obs}
\end{align}
 \end{subequations}
where ${\Phi}_{k+l,k}\eqg{A}_{k+l-1}\ldots{A}_{k}$, with  ${\Phi}_{k,k+l}={\Phi}_{k+l,k}^{-1}$, is the state transition matrix associated to ${A}_k\in\Rnn$ which is assumed to be \emph{invertible}; \linebreak $\bar{\lambda}_{k+l,k}\eqg{\lambda}_{k+l-1}\ldots{\lambda}_{k}$, where $\lambda_k$ is defined in \eqref{kf_A}; $V_k\in\R^{\bpk \times\bpk }$ is  an SPD matrix, ${R}_k\in\Rnm$ and $\bar{F}_{k}\in\R^{n\times\bpk }$.
\begin{defi}[uniform complete controllability]\label{defi_contr}
The matrix pair \linebreak
$\{\bar{A}_k,{R}_kW_k^{\frac{1}{2}}\}$ is \emph{uniformly completely controllable}, if there exist positive constants $\bar{\varrho}_1$ and $\bar{\varrho}_2$ and a positive integer ${h}$, such that, for all $k\geq h$, 
\mathc
\begin{align}
\bar{\varrho}_1I_n\leq \bar{\CC}_{k,k-{h}} \leq \bar{\varrho}_2I_n.
\end{align}
\end{defi}
\begin{defi}[uniform complete observability]\label{defi_obs}
The  matrix pair \linebreak $\{\bar{A}_{k},\bar{F}_{k}^T\}$ 
is \emph{uniformly completely observable}, if there exist positive constants $\bar{\varrho}_1$ and $\bar{\varrho}_2$ and a positive integer $h$, such that, for all $k\geq h$, 
\mathc
\begin{align}
\bar{\varrho}_1I_n\leq \bar{\OC}_{k,k-h} \leq \bar{\varrho}_2I_n.
\end{align}
\end{defi}
\begin{propo}[\cite{Son:95,Bag:09}]
Let $\bar{\KC}=\N^*$ (\cf\eqref{bp_KC_def}). If the matrix pairs  $\{\bar{A}_k,{R}_kW_k^{\frac{1}{2}}\}$ and $\{\bar{A}_{k},\bar{F}_{k}^T\}$ are uniformly completely   controllable and observable \resp., the estimation covariance matrix of the Kalman filter \eqref{kf}, designed for the system  \eqref{kf_sys}, satisfies the following inequalities, for all $k\geq l$:
\mathc
\begin{align}
(\bar{\OC}_{k,k-l}+\bar{\CC}_{k,k-l}^{-1})^{-1}\leq \bar{P}_k \leq \bar{\OC}_{k,k-l}^{-1}+\bar{\CC}_{k,k-l}.\nn
\end{align}
\end{propo}
\begin{propo}\label{proo_obs_con_unif}
The pairs $\{\bar{A}_k,{R}_kW_k^{\frac{1}{2}}\}$ and $\{\bar{A}_{k},\bar{F}_{k}^T\}$ are uniformly completely controllable and observable \resp., \ssi  $\{{A}_k,{R}_k\}$ and $\{{A}_{k},\bar{F}_{k}^T\}$ have the respective properties.
\end{propo}
\debdemo 
Since $\lambda_k$ and $W_k^{\frac{1}{2}}$, given in \eqref{kf_A} and \eqref{kf_W}, are a both bounded and positive  (\resp. SPD), the observability and controllability gramians, associated to the matrices $A_k$, $R_k$ and $\bar{F}_{k}$:
 ${\OC}_{k,k-l}\eqg\sum_{i=k}^{k+l}{\Phi}_{i,k}^T\bar{F}_{i}V_i^{-1}\bar{F}_{i}^T{\Phi}_{i,k}$  and ${\CC}_{k,k-l}\eqg\sum_{i=k}^{k+l-1}{\Phi}_{k,i+1}{R}_i{R}_i^T{\Phi}_{k,i+1}^T$, 
 are SPD bounded matrices  \ssi $\bar{\OC}_{k,k-l}$ and $\bar{\CC}_{k,k-l}$, given by \eqref{gramian_con_obs}, associated to $\lambda_kA_k$, $R_kW_k^{\frac{1}{2}}$ and $\bar{F}_{k}$ are also bounded SPD matrices.
\findemo 

It is needless to say that it is difficult to ensure the full rank for the matrix sum $\bar{\OC}_{k,k-{h}}$ \eqref{gramian_obs} on a time window of constant length $h$ when dealing with sporadic measurements. The system \eqref{system0}-\eqref{bounds} can therefore not be uniformly completely observable. Let us then introduce a new observability criterion for this kind of systems by using the observabilty gramian with a variable length:
\begin{defi}[sporadic observability]\label{defi_obs_spora} 
Assuming that $\bar{F}_{k}\eqg 0_{n\times \bpk }$, $\forall k\notin\bar{\KC}$ (\cf\eqref{bp_KC_def}), the  pair $\{\bar{A}_{k},\bar{F}_{k}^T\}$  
is said \emph{sporadically observable}, if there exist positive constants $\bar{\varrho}_1$ and $\bar{\varrho}_2$ and a positive integer $h$, such that, for all $k\geq \kappa_k(h)$,
\mathc
\begin{align*}
\bar{\varrho}_1I_n\leq \bar{\OC}_{k,k-\kappa_k(h)} \leq \bar{\varrho}_2I_n,
\end{align*}
where $\kappa_k(h)\in\N$  is \tq 
\begin{align}
\card\big(\{i\in\bar{\KC}| k-\kappa_k(h)\leq i \leq k\}\big)=h\label{card}
\end{align}
and where $\card(\SC)$ stands for the cardinality (number of elements) of the set $\SC$.
\end{defi}
The direct consequence of Proposition \ref{proo_obs_con_unif} applied to the system with all $i\in\gscrk\cap\dscrk\cap\hscrk$ measurements \eqref{bounds} including equality constraints \eqref{bound_HyperP0} can be stated as follows:
\begin{coro}\label{Lem_Bounds_Pbar}
Consider the system \eqref{system0} with \eqref{bounds} and the matrix $P_k$ computed in line with \eqref{predic_eq}, \eqref{predic_eq_glo} and \eqref{ellips_correc}. 
Let $H_{k_1}\in~\R^{n\times\bar{n}_k}$ 
whose columns form orthonormal basis for $\range \big(P_k\big)$ 
where $\bar{n}_k\eqg\rank(P_k)$ and let 
$\bar{P}_{k}\eqg H_{k_1}P_kH_{k_1}^T$. 
If the pair $\{A_{k},\bar{F}_{k}^T\}$ is sporadically observable and $\{A_{k},R_k\}$ is completely uniformly controllable,
then there exist positive finite numbers $\varrho_{k_1}$ and $\varrho_{k_2}$, \tq, for all $k\geq \kappa_k(l)$,
\mathc
\begin{align}
\varrho_{k_1}I_{\bar{n}_k}&\leq \bar{P}_k \leq \varrho_{k_2}I_{\bar{n}_k},\label{Bounds_Pbar}
\end{align}
$\kappa_k$ being defined in \eqref{card}.
\end{coro}
\subsection{Input-to-State stability}\label{sec_ISS}
In this paragraph, we shall examine the ISS concept. Before doing so, let us recall some comparison functions, widely used in stability analysis. A continuous function
$\psi_1:\R_+\rightarrow\R_+$ is called positive definite if it satisfies $\psi_1(0)=0$ and $\psi_1(t)>0$, $\forall t>0$. A positive definite function is of class $\mathscr{K}$ if it is strictly increasing and of class $\kf_{\infty}$ if it is of class $\mathscr{K}$ and unbounded. A continuous function $\psi_2:\R_+\rightarrow\R_+$ is of class  $\mathscr{L}$ if $\psi_2(t)$ is strictly decreasing to 0 as $t\rightarrow \infty$ and a continuous function $\psi_3:\R_+\times\R_+\rightarrow\R_+$ is of class $\mathscr{KL}$ if it is of class $\mathscr{K}$ in the first argument and of class $\mathscr{L}$ in the second argument. 
\begin{defi}[\bf\cite{Jia:01}]\label{defi_ISS}
The system 
\mathl
\begin{subequations}\label{ISS_sys}
\begin{align}
\zb(k+1)&=f(\zb(k),\ub(k)),\\ 
\text{where } \qquad f(\zero,\zero)&=\zero \text{ and }  \zb(0)\eqg\zbz, 
\end{align}
\end{subequations}
is globally input-to-state stable (ISS), if there exists a $\mathscr{KL}$-function $\beta$ and a $\mathscr{K}$-function $\psi$ such that, for each bounded input sequence $\ub_{[0,k]}\eqg\left\{\ub_0,\ldots,\ub_k\right\}$ and each $\zbz\in~\Rn$,
\mathc
\begin{align*}
\norm{\zb(k,\zbz,\ub_{[0,k-1]})}\leq\beta(\norm{\xza},k)+\psi({\sup_{i\leq k-1}\norm{\ub_{i}}}),
\end{align*}
where $\zb(k,\zbz,\ub_{[0,k-1]})$ is the trajectory of the system \eqref{ISS_sys}, for the initial state $\zbz\in\Rn$ and the input sequence $\ub_{[0,k-1]}$.
\end{defi}
\begin{defi}[\bf\cite{Jia:01}]\label{defi_ISS_Lyap}\emph{
A continuous function $\Vscr: \Rn\rightarrow\R_+$ is an ISS-Lyapunov function for the system \eqref{ISS_sys}, if there exists $\kf_{\infty}$-functions
    $\psi_1$
    and $\psi_2$ such that for all $\zb\in\Rn$, 
    $\psi_1(\norm{\zb})\leq\Vscr(\zb)\leq \psi_2(\norm{\zb})$
    and there exists a $\kf_{\infty}$-function $\psi_3$ and a
    $\mathscr{K}$-function $\chi$ such that for all $\zb\in\Rn$ and
    all $\ub\in\Rm$,
    $\Vscr\big(f(\zb,\ub)\big)-\Vscr(\zb)\leq
    -\psi_3(\norm{\zb})+\chi(\norm{\ub})$.  }
\end{defi}
\begin{lemm}[\bf\cite{Jia:01}]\label{lemm_ISS}
    The system \eqref{ISS_sys} 
    is ISS if it admits a continuous ISS-Lyapunov function. 
\end{lemm}
\subsection{Stability of the proposed estimation algorithm}\label{sec_thm_pf}
This section is dedicated to the proof  of  Theorem \ref{theo_ISS}.
\begin{enumerate}
\item Let $\underb{n}_k\eqg  n-\bar{n}_k=n-\rank(n)$ and $H_{k_2}\in~\R^{n\times \underb{n}_k}$ 
whose columns form orthonormal basis for $\nul \big(P_k\big)$,  \tq, $H_k\eqg\left[\left.H_{k_1}\right\vert H_{k_2}\right]$ is a unitary matrix. Hence, we have 
$H_k^TP_kH_k=\bdiag(\bar{P}_k,0_{\underb{n}_k\times\underb{n}_k})$. 
The $\EC_k$'s semi-axes lengths are the singular values of the matrix $\sigk P_k$, which are those of $\sigk H_k^TP_kH_k=\sigk \bar{P}_k$. On one hand, it is shown, at  the point \ref{theo_prop_lim_sig} of Thm \ref{theo_prop}, that the sequence $\big(\sigk\big)_{k\in\N}$ is decreasing, bounded above by $\sigz$ and convergent. On the other hand, as stated in \eqref{Bounds_Pbar} of Corollary \ref{Lem_Bounds_Pbar}, the singular values of $\bar{P}_k$ are bounded and so are the ellipsoid's axes lengths, as well as their product representing the ellipsoid's volume.
\item First, for any possible value of  the true state vector $\xka$, we have
\begin{align}
\xka\in\EC(\xk,\sigk P_k)&\eq \dxk\eqg\xka-\xk\in\EC(\zero_n,\sigk P_k)\nn\\
&\eq\dxk=(\sigki P_k)^{\frac{1}{2}}\ubk, \ubk\in\BC_2^n,\text{ (\cf $\S$ \ref{sec_intro}. \ref{unit_ball}).}
\end{align}
It means that $\dxk\in\range \big(P_k\big)$, 
which is a subspace of $\Rn$ of dimension $\bar{n}_k\leq n$: 
\begin{align}
H_k^T\dxk &=\sigk^{\frac{1}{2}}H_k^TP_k^{\frac{1}{2}}H_kH_k^T\ubk=\sigk^{\frac{1}{2}}\Big[{\bar{P}_k^{\frac{1}{2}}\atop 0_{ \underb{n}_k\times \underb{n}_k}} {0_{ \underb{n}_k\times \underb{n}_k}\atop 0_{ \underb{n}_k\times \underb{n}_k}}\Big]H_k^T\ubk
=\Big[{\sigk^{\frac{1}{2}}\bar{P}_k^{\frac{1}{2}}\bar{\ub}_k\atop \zero_{ \underb{n}_k}}\Big], \nn
\end{align}
where  $\bar{\ub}_k\eqg H_{k_1}^T\ubk\in\BC_2^{\bar{n}}$, meaning that 
\begin{align}
\forall\xka\in\EC_k, \dxk=H_k^T[\dxbki{1}^T \zero_{ \underb{n}_k}^T]^T, \text{ where }
\dxbki{1}\eqg H_{k_1}^T\dxbk \text{ and } H_{k_2}^T\dxbk =\zero_{ \underb{n}_k}.\nn
\end{align} 
%
Now we shall show that $\Vscr_k$ is an ISS-Lyapunov function for all possible values of $\dxk\in\range \big(P_k\big)$. 
First,   
\mathl
\begin{align}
\Vscr_k&\eqg \dxbk^TH_k H_k^TP_k^\dag H_k H_k^T\dxbk=\big[\dxbki{1}^T\vert\dxbki{2}^T\big] \Big[{\bar{P}_k^{-1}\atop 0} {0\atop 0}\Big] \big[\dxbki{1}^T\vert\dxbki{2}^T\big]^T=\dxbki{1}^T \bar{P}_k^{-1} \dxbki{1}\nn
\end{align}
noticing that   $\norm{\dxbki{1}}=\norm{H_k^T\dxk}=\norm{\dxk}$ and by virtue of \eqref{Bounds_Pbar}, it can be deduced that 
\mathc
\begin{align}
\psi_{k_2}(\norm{\dxk})&\leq\Vscr_k\leq\psi_{k_1}(\norm{\dxk}), 
\end{align} 
\mathl
\begin{align}
\text{ where }&&{
\begin{array}{rrcl}
\psi\ki:&\R_+&\rightarrow&\R_+,\quad i\in\{1,2\}, \\
&t &\mapsto&\psi\ki(t)=\varrho\ki^{-1}t^2,
\end{array}} \text{ are $\kf_{\infty}$ functions.}
\end{align} 
Second, 
from the point \ref{theo_prop_lim_sig} of Thm \ref{theo_prop}, we have 
\mathl
\begin{align}
&&\Vscr_k-\Vscr\kk&\leq -\sum_{i=0}^{p_k}{\alphak{i}\betak{i}^2\delkis}\leq0\label{theo_ISS_pf_1}\\
& \text{where } &\Vscr\kk&\eqg\dxkk^T\Pkk^{\dag}\dxkk \\
&\text{and where } &\dxkk&\eqg A\kkk\dxkkk+\etabkk
\end{align}
%
    Basing on the same reasoning as done in Lemma 3 in \cite{She:18}, it can be
    shown that for any vectors $\ab,\bb\in\Rn$ and any matrices $A,B\in\Rnn$, 
     \mathc
    \begin{align}
(\ab+\bb)^T(A+B)^{\dag}(\ab+\bb)\leq \ab^TA^{\dag}\ab+\bb^TB^{\dag}\bb
    \end{align}
    and considering $P\kkp$ given by \eqref{predic_eq_glo}, we have
    \mathl
    \begin{align}
    \Vscr\kk&\leq\tfrac{\bar{\mu}_{k-1_0}}{\bar{\mu}_{k-1_0}+\bar{\mu}_{k-1}}\Big(\dxkkk^TA\kkk^T\big(A\kkk\Pkkk A\kkk^T\big)^{\dag}A\kkk\dxkkk\nn\\
    &\pushr{+\tfrac{\sigkkk}{\bar{\mu}_{k-1_0}}\etabkk^T\bar{R}_{k-1}^{\dag}\etabkk\Big)}\nn\\
&\leq \tfrac{\bar{\mu}_{k-1_0}}{\bar{\mu}_{k-1_0}+\bar{\mu}_{k-1}}\dxkkk^T\Pkkk^{\dag}\dxkkk+\tfrac{\sigkkk}{\bar{\mu}_{k-1_0}+\bar{\mu}_{k-1}}\norm{\bar{R}_{k-1}^{\dag}}\norm{\etabkk}^2.\label{theo_ISS_pf_2}
    \end{align}
    Now, from \eqref{theo_ISS_pf_1}, we have
    \mathc
    \begin{align}
    \Vscr_k-\Vscr\kkk\leq\Vscr\kk-\Vscr\kkk;
    \end{align}
 and consequently, \eqref{theo_ISS_pf_2} becomes 
    \begin{align}
    \Vscr_k-\Vscr\kkk\leq\Vscr\kk-\Vscr\kkk\leq-\varrho_k\Vscr\kkk+\psi_k(\norm{\etabkk}),
    \end{align}
    \mathl
 where $\varrho_k\eqg\frac{\bar{\mu}_{k}}{\bar{\mu}_{k_0}+\bar{\mu}_{k}}>0 $
 and 
$\phi_k:\R_+\rightarrow\R_+$, 
$t \mapsto\phi_k(t)=\frac{\sigk\sigma_{k}}{\bar{\mu}_{k_0}+\bar{\mu}_{k}}t^2$,
 is a $\kf_{\infty}$ function,
where $\sigma_{k}=\norm{\big(R_k{M}_{k}^{-1}R_k^T\big)^{\dag}}>0$.
This means that $\Vscr_k$ is an ISS-Lyapunov function for the system of state vector $\dxk$. Thus
    applying Lemma \ref{lemm_ISS}  completes the proof of the theorem. 
    
    The cases where $k\notin\bar{\KC}$ can be viewed as measurements $i$ for which $\alphak{i}=0$ or $\betak{i}=0$.
    \end{enumerate}
\begin{rema}
$H_k\eqg\left[\left.H_{k_1}\right\vert H_{k_2}\right]\in\Rnn$ is a unitary matrix which rotates $P_k$ into a basis where it has two-bloc-diagonal form ($H_k$ can be obtained by QR decomposition of $P_k^{\frac{1}{2}}$ or by SVD of $P_k$).  $H_k^T\dxk=~[\dxbki{1}^T\vert\dxbki{2}^T]^T$ 
gives the components of  the state estimation error vector, $\dxbk$, in this new rotated basis, where the $\bar{n}_k$ first components are ISS (point \ref{theo_prop_ISS} of Thm \ref{theo_prop}) and the last $ \underb{n}_k$ ones are zero, 
meaning that the corresponding estimations are equal to their true values, in this rotated basis.
\end{rema}
{%
\bibliographystyle{alpha}
\bibliography{Biblio}
}
\end{document}